\RequirePackage{fix-cm}
\documentclass[smallextended]{svjour3}       % onecolumn (second format)
\smartqed  % flush right qed marks, e.g. at end of proof
\usepackage{graphicx}
\usepackage[utf8]{inputenc}
\usepackage[T1]{fontenc}
\usepackage{lmodern}
\usepackage[UKenglish]{babel}
\usepackage{graphicx}
\usepackage{amssymb,amsmath,mathtools}
\usepackage{scalerel}
\usepackage{pifont}
\usepackage{wasysym}
\usepackage{nicefrac}
\usepackage{epstopdf}
\usepackage{url}
\usepackage{csquotes}
\usepackage{wasysym}
\usepackage{tabularx}
\usepackage{float}
\usepackage{placeins}
\usepackage{tikz}
\usetikzlibrary{shapes.misc,automata,positioning,arrows,arrows.meta}
\usepackage{textcomp}
\usepackage{xspace}
\usepackage{xcolor}
\usepackage{booktabs}
\usepackage{enumitem}
\usepackage{subcaption}
\usepackage{csquotes}
\usepackage{multirow}
\usepackage{microtype}
\usepackage{hyperref}
\usepackage{doi}

\usepackage{algorithm}
\usepackage{algorithmic}

\tikzset{state/.style={circle,
            rounded corners,
            draw=black,
            minimum height=2em,
            minimum width=2em,
            align=center,
            },
	     paredge/.style={draw=teal,ultra thick},
	     paredgee/.style={draw=cyan,ultra thick},	     
         parsquare/.style={draw=black,rectangle},
         parcircle/.style={draw=black,circle,minimum width=1.2em}
}

\tikzset{every edge/.append style={-Triangle}}

\definecolor{darkgreen}{rgb}{0.0, 0.6, 0.0}
\newcommand{\cmark}{\textcolor{darkgreen}{\ding{51}}}
\newcommand{\xmark}{\textcolor{orange}{\textbf{\clock}}}

\newcommand{\tool}[1]{{\textsf{\sc #1}}}
\newcommand{\memout}{\textsf{\sc mem}}
\newcommand{\timeout}{\textsf{\sc time}}
\newcommand{\errorout}{\textsf{\sc err}}

\newcommand{\toolname}{\tool{Strix}}
\newcommand{\ltlsynt}{\tool{ltlsynt}}

\newcommand{\bosy}{\tool{BoSy}}
\newcommand{\syntcomp}{\tool{Syntcomp}}
\newcommand{\SC}{\tool{Syntcomp2019}}
\newcommand{\SCet}{\tool{Syntcomp2018}}

\newcommand{\true}{{\ensuremath{\mathbf{tt}}}}
\newcommand{\false}{{\ensuremath{\mathbf{ff}}}}
\newcommand{\F}{{\ensuremath{\mathbf{F}}}}
\newcommand{\G}{{\ensuremath{\mathbf{G}}}}
\newcommand{\X}{{\ensuremath{\mathbf{X}}}}
\newcommand{\U}{{\ensuremath{\mathbf{U}}}}

\newcommand{\R}{{\ensuremath{\mathbf{R}}}}

\newcommand{\N}{{\ensuremath{\mathbb{N}}}}

\newcommand{\lang}{\mathcal{L}}

\newcommand{\cO}{{\ensuremath{\mathcal{O}}}}
\newcommand{\cA}{{\ensuremath{\mathcal{A}}}}
\newcommand{\cP}{{\ensuremath{A}}}

\newcommand{\parity}{\ensuremath{p}}

\newcommand{\even}{\ensuremath{0}}
\newcommand{\odd}{\ensuremath{1}}
\DeclareMathSymbol{\symunknown}{\mathalpha}{operators}{"3F}
\newcommand{\abs}[1]{\ensuremath{\left|{#1}\right|}}

\newcommand{\acceptance}{\tau}
\newcommand{\decompose}{\mathcal{T}}
\newcommand{\odecomp}{\ensuremath{\cO_{\decompose}}}
\newcommand{\oinit}{\ensuremath{\cO_{q_0}}}
\newcommand{\oparity}{\ensuremath{\cO_{\parity}}}
\newcommand{\osucc}{\ensuremath{\cO_{\delta}}}

\newcommand{\oexpl}{\ensuremath{\cO_{\textsf{expl}}}}
\newcommand{\owin}{\ensuremath{\cO_{\textsf{win}}}}

\newcommand{\oscore}{\ensuremath{\cO_{\textsf{score}}}}

\newcommand{\oexplbfs}{\ensuremath{\cO_{\textsf{expl}}^{\textsf{bfs}}}}
\newcommand{\oexplbfsclear}{\ensuremath{\cO_{\textsf{expl}}^{\textsf{bfs+}}}}
\newcommand{\oexplpq}{\ensuremath{\cO_{\textsf{expl}}^{\textsf{pq}}}}
\newcommand{\oexplpqclear}{\ensuremath{\cO_{\textsf{expl}}^{\textsf{pq+}}}}
\newcommand{\dist}{\ensuremath{\textsf{dist}}}
\newcommand{\filter}{\ensuremath{\textsf{filter}}}
\newcommand{\winner}{\ensuremath{\textsf{winner}}}
\newcommand{\score}{\ensuremath{\textsf{score}}}

\newcommand{\lunstruct}{\ensuremath{l_{\textsf{unstr}}}}
\newcommand{\lstruct}{\ensuremath{l_{\textsf{struct}}}}

\DeclareMathOperator*{\argmax}{arg\,max}
\DeclareMathOperator*{\argmin}{arg\,min}

\newcommand{\ap}{\ensuremath{\mathsf{Ap}}}
\newcommand{\apin}{\ensuremath{{\ap_{\textsf{in}}}}}
\newcommand{\apout}{\ensuremath{{\ap_{\textsf{out}}}}}
\newcommand{\al}{\ensuremath{\Sigma}}
\newcommand{\alin}{\ensuremath{\al_{\textsf{in}}}}
\newcommand{\alout}{\ensuremath{\al_{\textsf{out}}}}

\newcommand{\plr}{\ensuremath{\mathsf{P}}}
\newcommand{\opp}{\overline{\plr}}
\newcommand{\symenv}{\mathbin{\square}}
\newcommand{\symctrl}{\mathord{\scalerel*{\bigcirc}{\square}}}
\newcommand{\pctrl}{\ensuremath{{\symctrl}}}
\newcommand{\penv}{\ensuremath{{\symenv}}}

\newcommand{\vctrl}{\ensuremath{V_{\symctrl}}}
\newcommand{\venv}{\ensuremath{V_{\symenv}}}

\newcommand{\pord}{\prec_{\plr}}

\journalname{Preprint}

\begin{document}

% Grants or other notes about the article that should go on the front
% page should be placed within the \thanks{} command in the title
% (and the %-sign in front of \thanks{} should be deleted)
%
% General acknowledgments should be placed at the end of the article.

\title{Practical Synthesis of Reactive Systems from LTL Specifications via Parity Games 
\thanks{This work was partially funded and supported by the German Research Foundation (DFG) projects \enquote{Game-based Synthesis for Industrial Automation} (253384115) and \enquote{Verified Model Checkers} (317422601) and the ERC Advanced Grant No. 787367 (PaVeS).}}

\titlerunning{Practical Synthesis of Reactive Systems from LTL Specifications} % if too long for running head
\subtitle{You \emph{can} teach an old dog new tricks: making a classic approach structured, forward-explorative, and incremental.}

\author{Michael Luttenberger \and Philipp J. Meyer \and Salomon Sickert}
\authorrunning{M. Luttenberger, P. J. Meyer, S. Sickert} % if too long for running head
\institute{Technical University of Munich, Germany \\
\email{\{luttenbe,meyerphi,sickert\}@in.tum.de} \\
All three authors contributed equally to this work and are listed in alphabetical order.
}


\maketitle

\begin{abstract}
The synthesis of reactive systems from linear temporal logic (LTL) specifications is an important aspect in the design of reliable software and hardware. 
We present our adaption of the classic automata-theoretic approach to LTL synthesis, %that we have
implemented in the tool \toolname{} which has won the two last synthesis competitions (\SCet{}/\tool{2019}).
The presented approach is (1) \emph{structured}, meaning that the states used in the construction have a semantic structure that is exploited in several ways, it performs a (2) \emph{forward exploration} such that it often constructs only a small subset of the reachable states, and it is (3) \emph{incremental} in the sense that it reuses results from previous inconclusive solution attempts. Further, we present and study different guiding heuristics that determine where to expand the on-demand constructed arena. Moreover, we show several techniques for extracting an implementation (Mealy machine or circuit) from the witness of the tree-automaton emptiness check. Lastly, the chosen constructions use a symbolic representation of the transition functions to reduce runtime and memory consumption. 
We evaluate the proposed techniques on the \SC{} benchmark set and show in more detail how the proposed techniques compare to the techniques implemented in other leading LTL synthesis tools.

\end{abstract}

\section{Introduction}\label{sec:introduction}

Synthesis refers to the problem of finding for a formal specification of an input-output relation a matching implementation \cite{Pnueli:1989}, e.g. an (I/O)-transducer, a Mealy machine, a Moore machine or a circuit. In our case we focus on \emph{linear temporal logic (LTL)} as the specification logic. While an asymptotically optimal  synthesis algorithm has been given in \cite{Pnueli:1989}, this approach and other algorithms solving this task\footnote{See \cite{DBLP:reference/mc/BloemCJ18} for an introduction to reactive synthesis and related graph games.} have not yet been successfully put into industrial practice. Tools able to deal with large specifications have been elusive and those that are available often produce subpar results compared to straight-forward manual implementations when successfully applied. \cite{DBLP:conf/sofsem/Kupferman12} identifies four challenges that hinder the practical impact of these synthesis algorithms: \enquote{algorithmic, methodological, scope, and qualitative} \cite{DBLP:conf/sofsem/Kupferman12}. The first challenge is to find efficient synthesis algorithms. The second challenge is the discrepancy between the assumed synthesis setting and reality: a finished and complete specification is the exception and not the rule. Users often iterate specifications and thus this brings up the task of reusing and composing intermediate results. The third challenge is expressiveness and succinctness of input and output formats. Finally, the fourth challenge is not only to compute any valid solution, but to find implementations that have \emph{good quality}. In this paper we primarily address the algorithmic and qualitative side of the synthesis problem, but also sketch ideas for the two other areas.

The classic automata-theoretic synthesis procedure using deterministic automata suffers from the \enquote{messy state space} \cite{DBLP:conf/sofsem/Kupferman12} of Safra's determinisation, which hinders efficient implementations that need to work on top of it. Moreover this automata-theoretic approach to synthesis requires the construction of a potentially double exponentially sized automaton (in the length of the specification)~\cite{DBLP:conf/dagstuhl/2001automata}. These two issues gave rise to \enquote{Safraless} approaches \cite{DBLP:conf/cav/KupfermanPV06,DBLP:conf/cav/FaymonvilleFT17,DBLP:conf/cav/BohyBFJR12,DBLP:conf/cav/KhalimovJB13,DBLP:conf/tacas/Ehlers11} to avoid the complicated state structure and to alleviate the state space explosion problem. Further, bounded synthesis adds to the synthesis problem an additional size constraint on the matching implementation. This effectively turns the synthesis problem into a search problem.

We, on the other hand, address the \enquote{messy state space} issue by employing a collection of \enquote{Safraless} LTL to \emph{deterministic parity automaton (DPA)} translations \cite{DBLP:conf/cav/SickertEJK16,DBLP:conf/tacas/EsparzaKRS17,DBLP:conf/lics/EsparzaKS18} in combination with a special product automaton construction that includes a \emph{latest appearance record (LAR)} construction and a formula decomposition in the spirit of \cite{DBLP:journals/corr/abs-1006-1408,DBLP:journals/fmsd/FiliotJR11,DBLP:journals/fmsd/Ehlers12,DBLP:journals/corr/abs-1709-02102}. Our construction recovers the Boolean structure present in the input specification. The state explosion problem is tackled by exploring the on-demand constructed parity game using a forward search resembling the optimisation described in \cite{DBLP:journals/ijfcs/FriedmannL12}. This enables our customised strategy iteration \cite{DBLP:journals/corr/abs-0806-2923} to leave most of the arena (and thus of the automaton) unexplored. Further, the decomposition allows us to split-off formulas and use them to prune the search-space which is a generalisation of a central insight from \cite{DBLP:journals/sttt/SohailS13}.

Further, we propose two heuristics guiding the construction of the arena in directions of probably decisive regions and thus focussing on important states, while skipping irrelevant parts. One approach is agnostic about the internal structure of the parity game, while the other one extracts information from the special parity automaton construction. Lastly, since we use strategy iteration to compute winning strategies, we can reuse so-far constructed solution attempts after expanding the arena, thereby reducing the amount of iterations until stabilisation is reached. We believe that our approach could be adapted to cache intermediate results (constructed automata, partial strategies) when using LTL synthesis interactively to speed up synthesis, which addresses the second area.

Regarding the quality of the synthesised structures \enquote{there is no {emphasize [sic]} on constructing optimal or well-structured systems} \cite{DBLP:conf/sofsem/Kupferman12}. While in this paper we do not look at general methods for producing qualitatively \emph{good} solutions and do not support synthesis under a specific quality measure, we provide a set of \emph{best-effort} heuristics to produce \emph{good} solutions: we make use of a range of post-processing steps to ensure that the solution is as small as possible. We also provide a modular encoding of the product automata into circuits that retains the Boolean structure of the specification. This approach surprisingly yields on some of the specifications used in the experimental evaluation smaller circuits compared to extracting a circuit out of a minimised Mealy machine. This data suggests that minimisation of the implementation represented as a Mealy machine might be in some cases diametral to generating small circuits. It seems that this area has not been studied enough and we think enriching specifications with additional, explicit optimisation goals is worthwhile, but currently these ideas have not manifested in specifications such as the \SC{} benchmarks.

We implement and test the outlined ideas within \toolname{}\footnote{\url{https://strix.model.in.tum.de/}} \cite{DBLP:conf/cav/MeyerSL18}, which relies on \cite{DBLP:conf/atva/KretinskyMS18} for automata translations and \cite{DBLP:conf/atva/MeyerL16} for parity game solving. An older version of \toolname{}  won in the TLSF/LTL track in all six categories of \SCet{} against other mature tools such as \ltlsynt{}~\cite{DBLP:journals/corr/abs-1711-11439}, which also implements synthesis using parity games, and \bosy{}~\cite{DBLP:conf/cav/FaymonvilleFT17}, which implements several bounded synthesis approaches. We further improve the prototype by representing the transition relation symbolically to address scalability issues for large alphabets and replace external tooling such as \tool{Speculoos} with an internal implementation able to cope with larger systems. This newer version again won in all TLSF/LTL tracks of \SC{} against \ltlsynt{}.

The rest of the paper is structured as follows: after introducing preliminaries, we give a high-level overview of the synthesis procedure and detail it in the following subsections. We then put our improvements to the test by evaluating them on the \SC{} benchmarks and comparing them with the old version, \ltlsynt{} and \bosy{}. Each section, if appropriate, contains its specific discussion of related work.

\paragraph{Editorial Note.} This paper is an extended version of the preliminary report published in~\cite{DBLP:conf/cav/MeyerSL18}. The synthesis approach is the same, but we give a much more detailed explanation of the techniques used, e.g., the decomposition and the product automaton construction. Further, we describe new extensions for different exploration strategies in Section~\ref{sec:boundary} and different encoding strategies in Section~\ref{sec:encoding}. We also give an updated experimental evaluation on a larger set of benchmarks and a comprehensive comparison with the old version and other tools.

\section{Preliminaries}

\subsection{$\omega$-Languages and $\omega$-Automata}
\label{subsec:automata}

Let $\Sigma$ be a finite alphabet. An $\omega$-word $w$ over $\Sigma$ is an infinite sequence of letters $a_0 a_1 a_2 \dots$ with $a_i \in \Sigma$ for all $i \geq 0$ and an $\omega$-language is a set of $\omega$-words. The set of all $\omega$-words is denoted $\Sigma^\omega$. We denote the $i$-th letter of an $\omega$-word $w$ (starting at 0) by $w(i)$ and the infinite suffix $w(i) w(i+1) \dots$ by $w_{i}$.

In this paper we focus on deterministic $\omega$-automata with accepting conditions defined on transitions which is nowadays the preferred acceptance condition in implementations due to the succinctness and in-line with other recent papers and tools \cite{DBLP:conf/forte/GiannakopoulouL02,DBLP:conf/cav/BabiakBDKKM0S15,DBLP:conf/atva/Duret-LutzLFMRX16,DBLP:conf/atva/KretinskyMS18}. The discussed constructions can also be transferred to automata with acceptance defined on states with the folklore translation from transition acceptance to state acceptance.

A \emph{deterministic pre-automaton (DA)} over $\Sigma$ is a tuple $(Q, \delta, q^0)$ where $Q$ is a finite set of states, $\delta \colon  Q \times \Sigma \rightarrow Q$ is a transition function, and $q^0$ is an initial state. A transition is a triple $(q, a, q')$ such that $q' = \delta(q, a)$.

A \emph{deterministic Parity automaton (DPA)} is a deterministic pre-automaton automaton $\cP = (Q, \delta, q^0, \chi, d, \parity)$ with the addition of the \emph{transition colouring} $\chi\colon Q\times \Sigma \to \{0,1,\ldots,d\}$, $d \geq 1$ the \emph{maximal colour} and $\parity \in \{ \even,\odd \}$ the \emph{parity} that determines whether a run is accepting or not (as defined below).
A {\em run} of $A$ on an $\omega$-word $w \colon \N_0 \rightarrow \Sigma$ is an $\omega$-sequence of states $\rho \colon \N_0 \rightarrow Q$ such that $\rho(0)=q^0$ and for all positions $i \in \N_0$, we have that $(\rho(i),w(i),\rho(i+1)) \in \delta$. Given a run $\rho$ over a word $w$, the infinite sequence of colours traversed by the run $\rho$ is denoted by $\chi(\rho)\coloneqq\bigl(\chi(\rho(i),w(i))\bigr)_{i\in\N_0}$.
The minimal colour appearing infinitely often along a run $\rho$ is $\liminf \chi(\rho)$.
A run $\rho$ is \emph{accepting} if $\liminf\chi(\rho) \equiv_2 \parity$ (with $x\equiv_2y :\Leftrightarrow (x-y)\bmod{2} = 0$).
An $\omega$-word $w$ is in the {\em language} of $\cP$, denoted $w \in \lang(\cP)$, iff the run for $w$ on $\cP$ is an accepting run.

For a parity $\parity \in \{\even,\odd\}$, define $\overline{\parity} \coloneqq 1-\parity$ as the \emph{switched parity}.
Note that by changing the parity $\parity$ to $\overline{\parity}$, we obtain a \emph{complement automaton}
$\overline{\cP} \coloneqq (Q, \delta, q^0, \chi, d, \overline{\parity})$ for which we have $\lang(\overline{\cP}) = \Sigma^\omega \setminus \lang(\cP)$.

To change the parity $\parity$ to $\overline{\parity}$ while preserving the language, one can use $\cP' \coloneqq (Q, \delta, q^0, \chi', d + 1, \overline{\parity})$ with $\chi'(q, a) \coloneqq \chi(q, a) + 1$, which has one more colour and satisfies $\lang(\cP') = \lang(\cP)$.

A \emph{deterministic B\"uchi automaton (DBA)} with the set of accepting transitions $\alpha$ is a DPA with colours $\{0,1\}$, parity $\even$ and $\chi$ defined as:
\[\chi(q, a) \coloneqq \begin{cases} 0 & \text{if } (q, a, \delta(q, a)) \in \alpha \\ 1 & \text{otherwise} \end{cases}\]

A \emph{deterministic co-B\"uchi automaton (DCA)} with the set of rejecting transitions $\beta$ is a DPA with colours $\{0,1\}$, parity $\odd$ and $\chi$ defined as:
\[\chi(q, a) \coloneqq \begin{cases} 0 & \text{if } (q, a, \delta(q, a)) \in \beta \\ 1 & \text{otherwise} \end{cases}\]

A \emph{deterministic weak automaton (DWA)} with the set of accepting states $\gamma$ is a DPA with colours $\{0,1\}$ and parity $\even$ or $\odd$, where for each strongly connected component $S \subseteq Q$, either $S \subseteq \gamma$ or $S \cap \gamma = \emptyset$.
Then $\chi$ is defined as:
\[\chi(q, a) \coloneqq
    \begin{cases}
        \parity & \text{if $\delta(q,a) \in \gamma$} \\
        \overline{\parity} & \text{otherwise} \\
    \end{cases}
\]
Note that for weak automata, we can switch between parity $\even$ and $\odd$ while preserving the language without increasing the number of colours.

A \emph{bottom state} of a DPA is a special state $\bot \in Q$ such that $\delta(\bot, a) = \bot$ and $\chi(\bot, a) \equiv_2 \overline{\parity}$ for each $a \in \Sigma$. A \emph{top state} of  a DPA is  a special state $\top \in Q$ such that $\delta(\top, a) = \top$ and $\chi(\top, a) \equiv_2 \parity$ for each $a \in \Sigma$.

\subsection{Linear Temporal Logic}

\newcommand{\subf}{\mathsf{sf}}
\newcommand{\ltlfum}{\mu{LTL}}
\newcommand{\ltlgrw}{\nu{LTL}}

We present LTL \cite{DBLP:conf/focs/Pnueli77} with a larger than usual set of modal operators and Boolean connectives, instead of a minimalistic syntax often found in other publications. While a minimalistic syntax reduces the amount of cases, e.g., in an induction, we want to keep as much structure of the given formula as possible and thus add redundancy. In particular, we are going to present customised constructions in order to deal with the Boolean connective $\leftrightarrow$ that effectively reduces the size of the automaton that is constructed. We work with a syntax for LTL in which formulas \emph{within} the scope of modal operators ($\X$, $\U$) are written in negation-normal form, i.e., negations only occur in front of atomic propositions. Thus we need to introduce $\false$, $\neg a$, $\vee$, and the temporal operator $\R$ (release) in order to remove $\neg$ from the syntax. It is easy to see that LTL formulas with the usual syntax can be translated to equivalent LTL formulas of the same size in our syntax. A formula of LTL over a set of atomic propositions ($\ap$) is given by the following syntax:

\begin{definition}[Syntax of LTL]
\begin{align*}
\varphi \Coloneqq \; & \varphi \wedge \varphi \mid \varphi \vee \varphi \mid \varphi \leftrightarrow \varphi \mid \psi \\
\psi \Coloneqq \; & \true \mid \false \mid a \mid \neg a \mid \psi \wedge \psi \mid \psi \vee \psi \mid \X\psi \; \mid \psi\U\psi \mid \psi\R\psi & \text{ with } a \in \ap
\end{align*}
\end{definition}%

We also define the usual abbreviations $\F \varphi \coloneqq \true \U \varphi$ (eventually) and $\G \varphi \coloneqq \false \R \varphi$ (always). The satisfaction relation $\models$ between $\omega$-words over the alphabet $\al \coloneqq 2^\ap$ and formulas is inductively defined as follows:
\begin{definition}[Semantics of LTL]
\[\begin{array}[t]{lclclcl}
w \models \true \\
w \not\models \false \\
w \models a & \mbox{ iff } & a \in w(0) \\
w \models \neg a & \mbox{ iff } & a \not \in w(0) \\
w \models \varphi \wedge \psi & \mbox{ iff } & w \models \varphi \text{ and } w \models \psi\\
w \models \varphi \vee \psi & \mbox{ iff } & w \models \varphi \text{ or } w \models \psi\\
w \models \varphi \leftrightarrow \psi & \mbox{ iff } & w \models \varphi \text{ if and only if } w \models \psi \\
w \models \X \varphi & \mbox{ iff } & w_1 \models \varphi\\
w \models \varphi\U \psi & \mbox{ iff } & \exists k. \, w_k \models \psi ~\text{ and }~ \forall j < k. \, w_j \models \varphi \\
w \models \varphi\R \psi & \mbox{ iff } & \forall k. \, w_k \models \psi ~\text{ or }~ \exists k. \, w_k \models \varphi ~\text{ and }~ \forall j \leq k. \, w_j \models \psi \\
\end{array}\]
We denote by $\lang(\varphi)$ the language of $\varphi$ defined as $\lang(\varphi) \coloneqq \{ w \in \al^\omega \mid w \models \varphi\}$.
\end{definition}%

\subsection{Notable Fragments of LTL}
\label{subsec:ltl-fragments} 

In the latter section we are going to consider the following four fragments of LTL:

\begin{itemize}
\item $\ltlfum$ and $\ltlgrw$:\\
$\ltlfum$ is the fragment of LTL restricted to the temporal operator $\U$, the Boolean connectives $(\wedge, \vee)$, the literals $(a, \neg a)$, and the next operator $(\X)$. $\ltlgrw$ is defined analogously, but with the operator $\R$ instead of $\U$. In the literature $\ltlfum$ is also called syntactic co-safety and $\ltlgrw$ syntactic safety.
\item $\G(\ltlfum)$ and $\F(\ltlgrw)$:\\
These fragments contain the formulas of the form $\G \varphi$, where $\varphi \in \ltlfum$, and $\F \varphi$, where $\varphi \in \ltlgrw$. 
\end{itemize}

The reason for the names $\ltlfum$ and $\ltlgrw$ is that $\U$ is a least-fixed-point operator, in the sense that its semantic is naturally formulated by a least fixed point, e.g., in the $\mu$-calculus, while the semantics of $\R$ is naturally formulated by a greatest fixed point.

For all these fragments several translations to deterministic automata are known and we are going to use the constructions for $\ltlfum$, $\ltlgrw$, $\G\F(\ltlfum)$, and $\F\G(\ltlgrw)$ described in \cite{DBLP:conf/lics/EsparzaKS18}, for $\G(\ltlfum)$ and $\F(\ltlgrw)$ the construction described in \cite{DBLP:conf/cav/SickertEJK16}, and for arbitrary LTL formulas the construction described in \cite{DBLP:conf/tacas/EsparzaKRS17}. It should be noted that of course these constructions can be swapped with other constructions, but some of the implemented heuristics rely on the specific state structure these constructions yield.

\subsection{Synthesis Problem}

Let $\varphi$ be a specification given as an LTL formula and let the atomic propositions $\ap = \apin \uplus \apout$ be partitioned into \emph{input symbols} $\apin$ and \emph{output symbols} $\apout$. We then define $\al \coloneqq 2^\ap$, $\alin \coloneqq 2^\apin$, and $\alout \coloneqq 2^\apout$.

Then the \emph{synthesis problem} is to decide if a function $\sigma : \alin^* \to \alout$ exists such that for every $\omega$-word $v \in \alin^\omega$, the $\omega$-word $w \in \al^\omega$ defined by $w(i) \coloneqq v(i) \cup \sigma(v(0)v(1)\ldots v(i))$ satisfies $w \in \lang(\varphi)$. In the positive case, also a finite and executable representation of $\sigma$ in the form of a controller should be produced, e.g., a Mealy machine or a circuit.

\section{Synthesis Procedure}\label{sec:algorithm}

We start with an overview on how \toolname{} constructs parity games from specification formulas and solves them. For the controller extraction we refer the reader to Section \ref{sec:extraction}. We illustrate the intuition of Algorithm~\ref{fig:synthesis-algorithm} using a simple arbiter example and refer to functionality explained in subsequent sections via \emph{oracles}: First, the formula is analysed with $\odecomp$ and a DPA is constructed on-the-fly via $\oinit$, $\oparity$, and $\osucc$. Second, the DPA is interpreted as a parity game and the game is solved via $\owin$ computing the winning regions. Further an exploration heuristic ($\oexpl$) guides which parts of the parity games are extended.  To be more precise, we use the following oracles:

\begin{algorithm}
\caption{Forward-explorative, incremental synthesis algorithm.}
\label{fig:synthesis-algorithm}
\begin{algorithmic}[1]
\REQUIRE LTL formula $\varphi$, input letters $\alin$, output letters $\alout$
\ENSURE $(\plr, \kappa)$, where $\plr$ is the winner of the game and $\kappa$ is a corresponding (non-deterministic) strategy.
\STATE $\alpha \leftarrow \odecomp(\varphi)$
\STATE $q_0 \leftarrow \oinit(\alpha)$
\STATE $(\vctrl, \venv) \leftarrow (\emptyset, \{q_0,\bot,\top\})$
\STATE $\parity \leftarrow \oparity(\alpha)$
\STATE $(E, \chi) \leftarrow (\{(\bot,\al,\bot),(\top,\al,\top)\}, \{((\bot,\al,\bot),\overline{\parity}),((\top,\al,\top),\parity)\})$
\STATE $\sigma, \tau \leftarrow (\emptyset,\emptyset)$
\STATE $B\leftarrow\{q_0\}$
\WHILE{$B \neq \emptyset$}
  \STATE $X \leftarrow \oexpl(\alpha, \vctrl, \venv, E, \chi, B, q_0, \parity, \sigma,\tau)$
  \STATE $B \leftarrow B \setminus X$
  \FORALL{$q \in X$}
     \FORALL{$(I,O,c,q') \in \osucc(\alpha,q)$}
       \IF{$q'\not\in \venv$}
       	 \STATE $B \leftarrow B \cup\{q'\}$
       \ENDIF
       \STATE $\vctrl \leftarrow \vctrl \cup \{(q, I)\}$
       \STATE $\venv  \leftarrow \venv  \cup \{q'\}$
       \STATE $E      \leftarrow E      \cup \{(q, I, (q, I)), ((q,I),O,q')\}$
       \STATE $\chi   \leftarrow \chi   \cup \{((q,I,(q,I)), \infty), (((q,I), O, q'), c)\}$
     \ENDFOR
  \ENDFOR
  \STATE $(\text{won}_{\symctrl}, \sigma) \leftarrow \owin(\vctrl, \venv, E, \chi, B, q_0, \pctrl, \parity, \sigma)$
  \IF{$\text{won}_{\symctrl}$}
    \RETURN $(\pctrl,\sigma)$
  \ENDIF
  \STATE $(\text{won}_{\symenv}, \tau) \leftarrow \owin(\vctrl, \venv, E, \chi, B, q_0, \penv, \overline{\parity}, \tau)$
  \IF{$\text{won}_{\symenv}$}
    \RETURN $(\penv,\tau)$
  \ENDIF
\ENDWHILE
\end{algorithmic}
\end{algorithm}

\begin{itemize}
  \item $\odecomp(\varphi)$: Given the formula $\varphi$, the oracle returns an annotated formula $\alpha$ that labels syntax nodes with a recommended automaton type to be used for this subformula for translation and how to compose a (product) DPA from these components.
  \item $\oinit(\alpha)$: Given the annotated formula $\alpha$, the oracle returns the initial state $q_0$ of the (product) DPA recognising $\varphi$.
  \item $\oparity(\alpha)$:  Given the annotated formula $\alpha$, the oracle returns the parity $\parity$ of the (product) DPA recognising $\varphi$.
  \item $\osucc(\alpha, q)$: Given the annotated formula $\alpha$ and a (product) state $q$,
      the oracle returns a set of outgoing transitions from $q$. The elements of the set are tuples $(I,O,c,q')$,
      where $I \subseteq \alin$ are the input letters, $O \subseteq \alout$ are the output letters, $c$ is the colour of the transition and $q'$ is the successor state. Formally for each such tuple $(I,O,c,q')$ we have $\delta(q, i \cup o) = q'$ and $\chi(q, i \cup o) = c$ for all $i \in I$ and $o \in O$.
      The oracle will return $q' = \bot$ for states that are trivially losing for the system and $q' = \top$ for states that are trivially winning.
  \item $\owin$: Given $(\vctrl, \venv, E, \chi, B, q, \plr, \parity, \kappa)$, compute whether the state $q$ is won by player $\plr\in\{\pctrl,\penv\}$,
  where $\vctrl$ is the set of nodes from which player $\pctrl$ moves, $\venv$ is the set of nodes from which player $\penv$ moves,
  $E\subseteq \left(\vctrl\cup\venv\right)\times2^\al\times\left(\vctrl\cup\venv\right)$ is the labeled edge relation, $\chi\colon E\to \N_0$ is an edge colouring, $B \subseteq \venv$ is the set of boundary nodes (i.e.\ nodes whose successors have yet to be constructed), $\parity$ is the parity for player $\plr$, i.e.\ player $\plr$ wins if the minimal colour occurring infinitely often along the edges of a play has parity $\parity$, and $\kappa$ is an initial (partial, nondeterministic)\footnote{We will later see the advantage of using nondeterministic strategies (multiple actions allowed) compared to deterministic strategies (only one action allowed).} strategy for player $\plr$. $\owin$ also returns an updated strategy $\kappa'$, which is winning from $q$ for $\plr$ if $q$ is won by player $\plr$. Depending on the player $\plr$ boundary nodes are declared as winning for the opponent in order to correctly under-approximate the parity game on the completely constructed arena.
  \item $\oexpl$: Given a set $B$ of boundary nodes, the so far constructed arena and the intermediate strategies, the oracle returns a nonempty subset of $B$ of nodes that should be further explored.
\end{itemize}

We use the specification of a simple arbiter as an example. In this setting two processes ($i\in\{1,2\}$) request access to the critical section by raising the flag $r_i$ and the arbiter eventually grants access to process $i$ by raising $g_i$. Thus we have $\apin = \{r_1,r_2\}$ and $\apout = \{g_1,g_2\}$ and the following specification:
\[
\phi =
\underbrace{\G (\neg g_1 \lor \neg g_2)}_{\psi_0} \land
\underbrace{\G (r_1 \rightarrow \F g_1)}_{\psi_1} \land
\underbrace{\G (r_2 \rightarrow \F g_2)}_{\psi_2}
\]

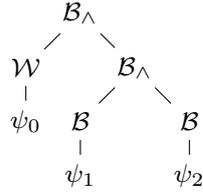
\begin{figure}
\begin{center}
\begin{tikzpicture}[sibling distance=4em, level distance=2em, scale=1.2,
    every node/.style = {align=center,scale=1.2}]
  \node {$\mathcal{B}_{\wedge}$}
   child { node {$\mathcal{W}$} 
           child { node {$\psi_0$} } 
         }
   child { node {$\mathcal{B}_{\wedge}$}
           child { node {$\mathcal{B}$} 
                   child { node {$\psi_1$} }
                 }
           child { node {$\mathcal{B}$}
                   child { node {$\psi_2$} }
                 }
         }
   ;
\end{tikzpicture}
\end{center}
    \caption{Annotated syntax tree $\alpha$ for $\phi$.}
    \label{fig:annotated-tree}
\end{figure}

\begin{figure}
    \centering
    \begin{subfigure}[b]{0.3\textwidth}
        \centering
        \begin{tikzpicture}[every node/.style={scale=0.75}]
    		\node[state,accepting] (q0) at (0,0) {$a_0$};
    		\node[state,right=4em of q0] (q1) at (0,0) {$\bot$};
    		\node (ghost) at (0,-0.5) {};
   			\path[->] (q0) edge [-{Latex[width=1.5mm,length=1mm]},loop left,looseness=4] node[left] {$\overline{g_1} + \overline{g_2},0$} (q0);
  			\path[->] (q0) edge node[-{Latex[width=1.5mm,length=1mm]},above] {$g_1 g_2,1$} (q1);
   			\path[->] (q1) edge [-{Latex[width=1.5mm,length=1mm]},loop right,looseness=4] node[right] {$\Sigma,1$} (q1);
  		\end{tikzpicture}
        \caption{$\mathcal{A}_{\psi_0}$: DWA, $\parity=0$}
    \end{subfigure}
    ~ %add desired spacing between images, e. g. ~, \quad, \qquad, \hfill etc. 
    \begin{subfigure}[b]{0.3\textwidth}
        \centering
        \begin{tikzpicture}[every node/.style={scale=0.75}]
    		\node[state] at (0,0) (q0) {$b_0$};
   			\node[state,right=2.5em of q0] (q1) {$b_1$};
   			\node (ghost) at (0,-0.5) {};
 		    \path[->] (q0) edge [double,-{Latex[width=1.5mm,length=1mm]},loop left,looseness=4] node[left] {$\overline{r_1},0$} (q0);
			\path[->,bend left=30] (q0) edge node[-{Latex[width=1.5mm,length=1mm]},above] {$r_1,1$} (q1);
			\path[->] (q1) edge [-{Latex[width=1.5mm,length=1mm]},loop right,looseness=4] node[right] {$\overline{g_1},1$} (q1);
			\path[->,bend left=30] (q1) edge[double,-{Latex[width=1.5mm,length=1mm]}] node[below] {$g_1,0$} (q0);
  		\end{tikzpicture}
        \caption{$\mathcal{A}_{\psi_1}$: DBA, $\parity=0$}
    \end{subfigure}
    ~ %add desired spacing between images, e. g. ~, \quad, \qquad, \hfill etc. 
    \begin{subfigure}[b]{0.3\textwidth}
    	\centering
        \begin{tikzpicture}[every node/.style={scale=0.75}]
			\node[state] at (0,0) (q0) {$c_0$};
			\node[state,right=2.5em of q0] (q1) {$c_1$};
			\node (ghost) at (0,-0.5) {};
			\path[->] (q0) edge [double,-{Latex[width=1.5mm,length=1mm]},loop left,looseness=4] node[left] {$\overline{r_2},0$} (q0);
			\path[->,bend left=30] (q0) edge[-{Latex[width=1.5mm,length=1mm]}] node[above] {$r_2,1$} (q1);
			\path[->] (q1) edge [-{Latex[width=1.5mm,length=1mm]},loop right,looseness=4] node[right] {$\overline{g_2},1$} (q1);
			\path[->,bend left=30] (q1) edge[double,-{Latex[width=1.5mm,length=1mm]}] node[below] {$g_2,0$} (q0);
		\end{tikzpicture}
        \caption{$\mathcal{A}_{\psi_2}$: DBA, $\parity=0$}
    \end{subfigure}
    \caption{DAs for $\psi_0$, $\psi_1$, and $\psi_2$. Note that $\psi_1$ and $\psi_2$ are isomorphic up to alphabet renaming.}
    \label{fig:automata}
\end{figure}
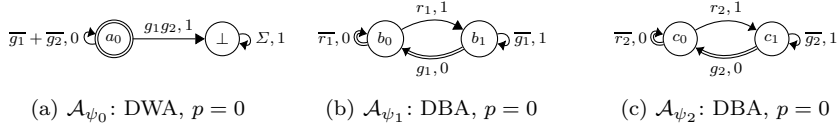

\noindent Applying the annotation oracle, we obtain from the specification $\phi$ an annotated syntax tree $\alpha$ shown in Fig.~\ref{fig:annotated-tree} that represents a decomposition of $\phi$ into subformulas w.r.t.\ the weakest class of deterministic automata needed for their translation and how to combine the automata in order to obtain the automaton for $\phi$ itself.
In our example, $\psi_0$ is a simple mutex requirement which is classified as recognisable by a DWA (denoted by $\mathcal{W}$), and $\psi_1$ and $\psi_2$ are fairness requirements that are classified as recognisable by a DBA (denoted by $\mathcal{B}$). The conjunctions $\psi_1 \wedge \psi_2$ and $\phi$ itself are then also recognisable by a DBA (denoted by $\mathcal{B}_{\wedge}$).
The corresponding automata are displayed in Fig.~\ref{fig:automata}.

We query $\oinit(\alpha)$ and obtain $q_0 \coloneqq (a_0,((b_0,c_0),0))$, which matches the tree structure of $\alpha$. Here, we have add a round-robin counter $r\in\{0,1\}$ for the intersection of the two Büchi automata representing $\psi_1$ and $\psi_2$: This round-robin counter remembers which of the two Büchi automata is due to take an accepting transition. To ease notation for our example and the corresponding figures we flatten $(a_0,((b_0,c_0),0))$ to $(a_0,\underline{b_0},c_0)$ with the underlined state representing  the round-robin counter. Further, we query $\oparity(\alpha)$ for the parity associated with the controller, i.e. player $\pctrl$. In our example it is $0$ and $\pctrl$ wins a play if the minimal colour encountered infinitely often is even.

As we are using Mealy semantics, we let the environment $\penv$ move from the initial node $(a_0,\underline{b_0},c_0)$. Our parity game also includes two nodes $\bot$ and $\top$ where by our construction $\bot$ is always won by the environment $\penv$, while $\top$ is always won by the controller $\pctrl$.

We now start the on-the-fly forward exploration of parity game arena. In every iteration of the while-loop in Algorithm~\ref{fig:synthesis-algorithm} we extend the \emph{boundary} $B$. The boundary always consists of nodes belonging to the environment $\penv$ whose successors have yet to be explored. Initially the boundary is just the initial node of the parity game, in our example $B=\{(a_0,\underline{b_0},c_0)\}$.

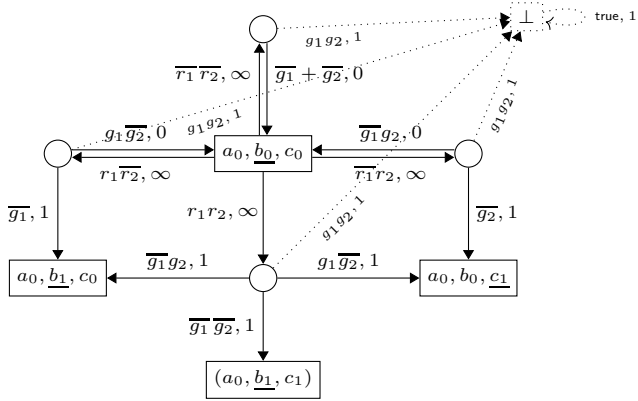
\begin{figure}[h]
\begin{center}
\begin{tikzpicture}[scale=1.5,font={\scriptsize}]
    \node[parsquare,draw] (q0) at (0,2) {$a_0,\underline{b_0},c_0$};
    \node[parcircle,draw] (q0i0) at (0,3.1) {};
    \node[parcircle,draw] (q0i1) at (-1.8,2) {};
    \node[parcircle,draw] (q0i2) at (1.8,2) {};
    \node[parcircle,draw] (q0i3) at (0,0.9) {};

    \node[parsquare,draw] (q1) at (-1.8,0.9) {$a_0,\underline{b_1},c_0$};
%    \node[parcircle,draw] (q1i0) at (-3.4,0.9) {};
%    \node[parcircle,draw] (q1i1) at (-1.8,0) {};
%    \node (q1i0) at (-3.4,0.9) {};
%    \node (q1i1) at (-1.8,0) {};

    \node[parsquare,draw] (q2) at (1.8,0.9) {$a_0,b_0,\underline{c_1}$};
%    \node[parcircle,draw] (q2i0) at (3.4,0.9) {};
%    \node[parcircle,draw] (q2i1) at (1.8,0) {};
%    \node (q2i0) at (3.4,0.9) {};
%    \node (q2i1) at (1.8,0) {};

    \node[parsquare,draw] (q3) at (0,0.0) {$(a_0,\underline{b_1},c_1)$};

    \path[->,transform canvas={xshift=-0.5mm}] (q0) edge node[left,yshift=2mm] {$\overline{r_1}\, \overline{r_2}, \infty$} (q0i0);
%    \path[->,paredge,transform canvas={xshift=0.5mm}] (q0i0) edge node[right,yshift=2mm] {$\overline{g_1} + \overline{g_2}$} (q0);
    \path[->,transform canvas={xshift=0.5mm}] (q0i0) edge node[right,yshift=2mm] {$\overline{g_1} + \overline{g_2},0$} (q0);
%    \path[->,paredge,transform canvas={yshift=0.5mm}] (q0i1) edge node[above,xshift=-1mm] {$g_1\overline{g_2}$} (q0);
    \path[->,transform canvas={yshift=-0.5mm}] (q0) edge node[below,xshift=-1mm] {$r_1 \overline{r_2}, \infty$} (q0i1);
    \path[->,transform canvas={yshift=0.5mm}] (q0i1) edge node[above,xshift=-1mm] {$g_1\overline{g_2},0$} (q0);
%    \path[->,dashed] (q0i1) edge node[left] {$\overline{g_1}$} (q1);
    \path[->] (q0i1) edge node[left] {$\overline{g_1},1$} (q1);
    \path[->,transform canvas={yshift=-0.5mm}] (q0) edge node[below,xshift=1mm] {$\overline{r_1} r_2, \infty$} (q0i2);
%    \path[->,paredge,transform canvas={yshift=0.5mm}] (q0i2) edge node[above,xshift=1mm] {$\overline{g_1}g_2$} (q0);
    \path[->,transform canvas={yshift=0.5mm}] (q0i2) edge node[above,xshift=1mm] {$\overline{g_1}g_2,0$} (q0);
%    \path[->,dashed] (q0i2) edge node[right] {$\overline{g_2}$} (q2);
    \path[->] (q0i2) edge node[right] {$\overline{g_2},1$} (q2);
    \path[->] (q0) edge node[left,xshift=0.5mm,yshift=0.5mm] {$r_1 r_2, \infty$} (q0i3);
%    \path[->,draw=orange,thick] (q0i3) edge node[above] {$\overline{g_1}g_2$} (q1);
    \path[->] (q0i3) edge node[above] {$\overline{g_1}g_2,1$} (q1);
%    \path[->,draw=orange,thick] (q0i3) edge node[above] {$g_1\overline{g_2}$} (q2);
    \path[->] (q0i3) edge node[above] {$g_1\overline{g_2},1$} (q2);
%    \draw[->,dashed] (q0i3) edge ([yshift=2mm]q3.north);
    \path[->] (q0i3) edge node[left,yshift=0mm] {$\overline{g_1}\,\overline{g_2},1$} (q3);

    \node[parsquare,draw,dotted] (bot) at (0,0.9,-6) {$\bot$};
    \path[dotted,->] (bot) edge[loop right] node[right=1pt] {\tiny $\textsf{true},1$} (bot);
    \draw[dotted,->] (q0i0) edge node[sloped,below,xshift=-8mm] {\tiny $g_1g_2,1$} (bot);
    \draw[dotted,->] (q0i1) edge node[sloped,below,xshift=-11mm] {\tiny $g_1g_2,1$} (bot);
    \draw[dotted,->] (q0i2) edge node[sloped,below,xshift=-2mm] {\tiny $g_1g_2,1$} (bot);
    \draw[dotted,->] (q0i3) edge node[sloped,below,xshift=-11mm] {\tiny $g_1g_2,1$} (bot);

%    \path[->,transform canvas={yshift=0.5mm}] (q1) edge node[above] {$\overline{r_1}$} (q1i0);
%    \path[->,dashed] (q1) edge (q1i0);
%    \path[->,paredge,bend left=55] (q1i0) edge node[above] {$g_0\overline{g_1}$} (q0);
%    \path[->,dashed,transform canvas={yshift=-0.5mm}] (q1i0) edge node[below] {$\overline{g_0}$} (q1);
%    \path[->,transform canvas={xshift=-0.5mm}] (q1) edge node[left] {$r_1$} (q1i1);
%    \path[->,dashed] (q1) edge  (q1i1);
%    \path[->,paredge] (q1i1) edge node[below,xshift=-4mm,yshift=-1mm] {$g_0\overline{g_1}$} (q2);
%    \path[->,dashed,transform canvas={xshift=0.5mm}] (q1i1) edge node[right] {$\overline{g_0}g_1$} (q1);
%    \draw[->,dashed,-Triangle] (q1i1) |- ([xshift=-2mm]q3.west);

%    \path[->,transform canvas={yshift=0.5mm}] (q2) edge node[above] {$\overline{r_0}$} (q2i0);
%    \path[->,dashed] (q2) edge (q2i0);
%    \path[->,paredge,bend right=55] (q2i0) edge node[above] {$\overline{g_0}g_1$} (q0);
%    \path[->,dashed,transform canvas={yshift=-0.5mm}] (q2i0) edge node[below] {$\overline{g_1}$} (q2);
%    \path[->,transform canvas={xshift=-0.5mm}] (q2) edge node[left] {$r_0$} (q2i1);
%	\path[->,dashed] (q2) edge (q2i1);
%    \path[->,paredge] (q2i1) edge node[below,xshift=4mm,yshift=-1mm] {$\overline{g_0}g_1$} (q1);
%    \path[->,dashed,transform canvas={xshift=0.5mm}] (q2i1) edge node[right] {$g_0\overline{g_1}$} (q2);
%    \draw[->,dashed,-Triangle] (q2i1) |- ([xshift=2mm]q3.east);
\end{tikzpicture}
\end{center}
\caption{Parity game arena after one iteration of the main loop.}\label{fig:pg-ex-1}
\end{figure}

As initially $B$ is a singleton set, $\oexpl$ tells us to explore all direct successors of $q_0=(a_0,\underline{b_0},c_0)$. $\osucc(\alpha, q_0)$ groups the outgoing transitions $\delta(q_0, \star)$ using $\alin$ and $\alout$ as previously mentioned as a set of tuples of the shape $(I, O, q', c)$. Due to the Mealy semantics each such tuple $(I, O, q', c)$ is broken up in two steps: starting in $q_0$, first the environment issues a signal $i \in I$, which leads the game into the intermediate state $(q_0,I)$;\footnote{We represent the intermediate states simply by circular shaped nodes in the figures.} as this is only an intermediate step the corresponding edge is assigned the (w.r.t.\ min-parity) \enquote{neutral} colour $\infty$; in $(q_0,I)$ the controller $\pctrl$ then issues a signal $o \in O$ leading to the next state of the DPA. We illustrate this using our example: we have $4$ choices for the inputs at $(a_0,\underline{b_0},c_0)$:
\begin{enumerate}
\item $\overline{r_1}\, \overline{r_2} = \{\emptyset\}$, i.e.\ no process requests access;
\item $r_1\, \overline{r_2} = \{\{r_1\}\}$, i.e.\ only process $1$ requests access;
\item $\overline{r_1}\, r_2 = \{\{r_2\}\}$, i.e.\ only process $2$ requests access; and
\item $r_1\, r_2 = \{\{r_1,r_2\}\}$, i.e.\ both processes want to access the critical section.
\end{enumerate}
Consider the case that the environment chooses the input $I=r_1\,\overline{r_2}$. $\osucc$ groups the outputs available to $\pctrl$ into the three groups $g_1\,g_2$ (grant access to both processes), $g_1\,\overline{g_2}$ (grant access to only process 1), and $\overline{g_1}$ (do not grant access to process 1). In case of $O=\overline{g_1}=\{\{\}, \{g_2\}\}$, the DBA $\cA_{\psi_1}$ takes a non-accepting transition, hence, the round-robin counter stays unchanged, while the other two automata take a loop, arriving at state $(a_0,\underline{b_1},c_0)$ in the product; $\osucc$ determines by analysing $\alpha$ that we only need the colours $\{0,1\}$ for the parity game under construction; as $\cA_{\psi_1}$ takes a non-accepting transition, $\osucc$ gives the input-output pair $(I, O) = (r_1\,\overline{r_2},\overline{g_1})$ a colour of parity $\overline{\parity}$ in order to prevent $\pctrl$ from replying to $r_1\,\overline{r_2}$ by  $\overline{g_1}$ infinitely often. Thus $(r_1\,\overline{r_2},\overline{g_1},1,(a_0,\underline{b_1},c_0)) \in \osucc(\alpha, q_0)$. For $O=g_1\,\overline{g_2}$ we obtain analogously the entry $(r_1\,\overline{r_2}, g_1\,\overline{g_2},0,(a_0,\underline{b_0},c_0))$: as all automata take an accepting loop in this case the input-output pair is given the colour $0$ and, as the round-robin counter is incremented twice, we are back in the initial node of the parity game. Finally, for $O=g_1\,g_2$ the oracle $\osucc$ determines that $\pctrl$ has no chance of winning anymore as the DWA $\cA_{\psi_0}$ representing the mutex requirements cannot reach an accepting transition anymore; for this reason, $\osucc$ simplifies the successor state to $\bot$ which by construction is always won by the environment $\penv$ and includes the tuple $(r_1\,\overline{r_2},g_1g_2,1,\bot)$ into its output.\footnote{Should $\osucc$ determine that all automata will accept any possible input-output pairs from now on, it simplifies the successor state to $\top$ which, again by construction, is always won by $\pctrl$.} Analogously, $\osucc$ handles the other three possible inputs by $\penv$ which eventually leads to the arena shown in Fig.~\ref{fig:pg-ex-1} with border $B=\{(a_0,\underline{b_0},c_1), (a_0,b_0,\underline{c_1}), (a_0,\underline{b_1},c_1)\}$.

\begin{figure}[h]
\begin{center}
\begin{tikzpicture}[scale=1.5,font=\scriptsize]
    \node[parsquare,draw] (q0) at (0,2) {$a_0,\underline{b_0},c_0$};
    \node[parcircle,draw] (q0i0) at (0,3.1) {};
    \node[parcircle,draw] (q0i1) at (-1.8,2) {};
    \node[parcircle,draw] (q0i2) at (1.8,2) {};
    \node[parcircle,draw] (q0i3) at (0,0.9) {};

    \node[parsquare,draw] (q1) at (-1.8,0.9) {$a_0,\underline{b_1},c_0$};
%    \node[parcircle,draw] (q1i0) at (-3.4,0.9) {};
%    \node[parcircle,draw] (q1i1) at (-1.8,0) {};
%    \node (q1i0) at (-3.4,0.9) {};
%    \node (q1i1) at (-1.8,0) {};

    \node[parsquare,draw] (q2) at (1.8,0.9) {$a_0,b_0,\underline{c_1}$};
%    \node[parcircle,draw] (q2i0) at (3.4,0.9) {};
%    \node[parcircle,draw] (q2i1) at (1.8,0) {};
%    \node (q2i0) at (3.4,0.9) {};
%    \node (q2i1) at (1.8,0) {};

    \node[parsquare,draw] (q3) at (0,0.0) {$(a_0,\underline{b_1},c_1)$};

    \path[->,transform canvas={xshift=-0.5mm}] (q0) edge node[left,yshift=2mm] {$\overline{r_1}\, \overline{r_2}, \infty$} (q0i0);
    \path[->,paredge,transform canvas={xshift=0.5mm}] (q0i0) edge node[right,yshift=2mm] {$\overline{g_1} + \overline{g_2}, 0$} (q0);
%    \path[->,transform canvas={xshift=0.5mm}] (q0i0) edge node[right,yshift=2mm] {$\overline{g_1} + \overline{g_2}$} (q0);
    \path[->,paredge,transform canvas={yshift=0.5mm}] (q0i1) edge node[above,xshift=-1mm] {$g_1\overline{g_2}, 0$} (q0);
    \path[->,transform canvas={yshift=-0.5mm}] (q0) edge node[below,xshift=-1mm] {$r_1 \overline{r_2}, \infty$} (q0i1);
%    \path[->,transform canvas={yshift=0.5mm}] (q0i1) edge node[above,xshift=-1mm] {$g_1\overline{g_2}$} (q0);
    \path[->,dotted] (q0i1) edge node[left] {$\overline{g_1},1$} (q1);
%    \path[->] (q0i1) edge node[left] {$\overline{g_1}$} (q1);
    \path[->,transform canvas={yshift=-0.5mm}] (q0) edge node[below,xshift=1mm] {$\overline{r_1} r_2, \infty$} (q0i2);
    \path[->,paredge,transform canvas={yshift=0.5mm}] (q0i2) edge node[above,xshift=1mm] {$\overline{g_1}g_2, 0$} (q0);
%    \path[->,transform canvas={yshift=0.5mm}] (q0i2) edge node[above,xshift=1mm] {$\overline{g_1}g_2$} (q0);
    \path[->,dotted] (q0i2) edge node[right] {$\overline{g_2}, 1$} (q2);
%    \path[->] (q0i2) edge node[right] {$\overline{g_2}$} (q2);
    \path[->] (q0) edge node[left,xshift=0.5mm,yshift=0.5mm] {$r_1 r_2, \infty$} (q0i3);
    \path[->, dotted] (q0i3) edge node[above] {$\overline{g_1}g_2, 1$} (q1);
%    \path[->] (q0i3) edge node[above] {$\overline{g_1}g_2$} (q1);
    \path[->, dotted] (q0i3) edge node[above] {$g_1\overline{g_2}, 1$} (q2);
%    \path[->] (q0i3) edge node[above] {$g_1\overline{g_2}$} (q2);
%    \draw[->,dotted] (q0i3) edge ([yshift=2mm]q3.north);
    \path[->, dotted] (q0i3) edge node[left,yshift=0mm] {$\overline{g_1}\,\overline{g_2}, 1$} (q3);

\end{tikzpicture}
\end{center}
\caption{Optimal strategy for the controller in the parity game arena of Fig.~\ref{fig:pg-ex-1}.}\label{fig:pg-ex-2}
\end{figure}
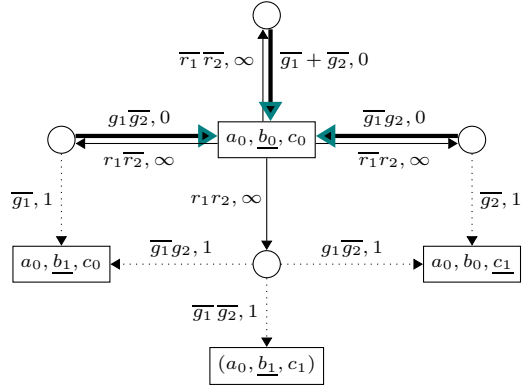

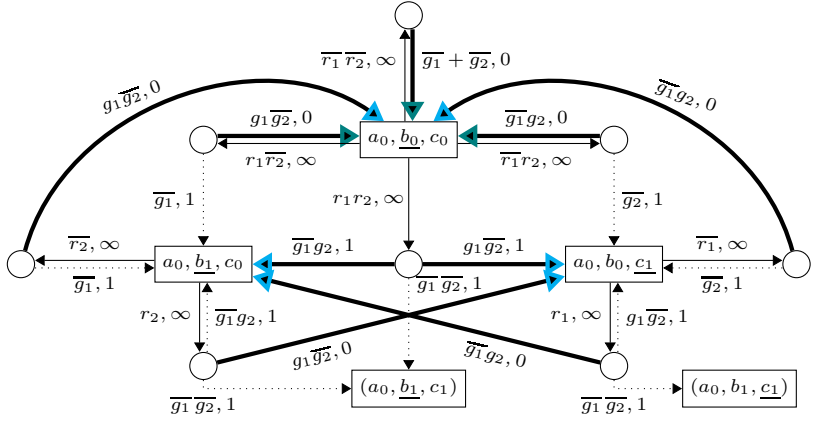
\begin{figure}[h]
\begin{center}
\begin{tikzpicture}[scale=1.5,font=\scriptsize]
    \node[parsquare,draw] (q0) at (0,2) {$a_0,\underline{b_0},c_0$};
    \node[parcircle,draw] (q0i0) at (0,3.1) {};
    \node[parcircle,draw] (q0i1) at (-1.8,2) {};
    \node[parcircle,draw] (q0i2) at (1.8,2) {};
    \node[parcircle,draw] (q0i3) at (0,0.9) {};

    \node[parsquare,draw] (q1) at (-1.8,0.9) {$a_0,\underline{b_1},c_0$};
    \node[parcircle,draw] (q1i0) at (-3.4,0.9) {};
    \node[parcircle,draw] (q1i1) at (-1.8,0) {};

    \node[parsquare,draw] (q2) at (1.8,0.9) {$a_0,b_0,\underline{c_1}$};
    \node[parcircle,draw] (q2i0) at (3.4,0.9) {};
    \node[parcircle,draw] (q2i1) at (1.8,0) {};

%    \node (q3) at (0,-0.3) {$\cdots$};
    \node[parsquare,draw] (q3) at (0,-0.2) {$(a_0,\underline{b_1},c_1)$};

    \node[parsquare,draw] (q4) at (2.9,-0.2) {$(a_0,b_1,\underline{c_1})$};

        \path[->,transform canvas={xshift=-0.5mm}] (q0) edge node[left,yshift=2mm] {$\overline{r_1}\, \overline{r_2}, \infty$} (q0i0);
        \path[->,paredge,transform canvas={xshift=0.5mm}] (q0i0) edge node[right,yshift=2mm] {$\overline{g_1} + \overline{g_2}, 0$} (q0);
    %    \path[->,transform canvas={xshift=0.5mm}] (q0i0) edge node[right,yshift=2mm] {$\overline{g_1} + \overline{g_2}$} (q0);
        \path[->,paredge,transform canvas={yshift=0.5mm}] (q0i1) edge node[above,xshift=-1mm] {$g_1\overline{g_2}, 0$} (q0);
        \path[->,transform canvas={yshift=-0.5mm}] (q0) edge node[below,xshift=-1mm] {$r_1 \overline{r_2}, \infty$} (q0i1);
    %    \path[->,transform canvas={yshift=0.5mm}] (q0i1) edge node[above,xshift=-1mm] {$g_1\overline{g_2}$} (q0);
        \path[->,dotted] (q0i1) edge node[left] {$\overline{g_1},1$} (q1);
    %    \path[->] (q0i1) edge node[left] {$\overline{g_1}$} (q1);
        \path[->,transform canvas={yshift=-0.5mm}] (q0) edge node[below,xshift=1mm] {$\overline{r_1} r_2, \infty$} (q0i2);
        \path[->,paredge,transform canvas={yshift=0.5mm}] (q0i2) edge node[above,xshift=1mm] {$\overline{g_1}g_2, 0$} (q0);
    %    \path[->,transform canvas={yshift=0.5mm}] (q0i2) edge node[above,xshift=1mm] {$\overline{g_1}g_2$} (q0);
        \path[->,dotted] (q0i2) edge node[right] {$\overline{g_2}, 1$} (q2);
    %    \path[->] (q0i2) edge node[right] {$\overline{g_2}$} (q2);
        \path[->] (q0) edge node[left,xshift=0.5mm,yshift=0.5mm] {$r_1 r_2, \infty$} (q0i3);
        \path[->, paredgee] (q0i3) edge node[above] {$\overline{g_1}g_2, 1$} (q1);
    %    \path[->] (q0i3) edge node[above] {$\overline{g_1}g_2$} (q1);
        \path[->, paredgee] (q0i3) edge node[above] {$g_1\overline{g_2}, 1$} (q2);
    %    \path[->] (q0i3) edge node[above] {$g_1\overline{g_2}$} (q2);
    %    \draw[->,dotted] (q0i3) edge ([yshift=2mm]q3.north);
        \path[->, dotted] (q0i3) edge node[right,yshift=5.3mm] {$\overline{g_1}\,\overline{g_2}, 1$} (q3);

    \path[->,transform canvas={yshift=0.5mm}] (q1) edge node[above] {$\overline{r_2}, \infty$} (q1i0);
    \path[->,paredgee,bend left=55] (q1i0) edge node[above,sloped,xshift=-5mm] {$g_1\overline{g_2}, 0$} (q0);
    \path[->,dotted,transform canvas={yshift=-0.5mm}] (q1i0) edge node[below] {$\overline{g_1}, 1$} (q1);
    \path[->,transform canvas={xshift=-0.5mm}] (q1) edge node[left] {$r_2, \infty$} (q1i1);
    \path[->,paredgee] (q1i1) edge node[below,xshift=-10mm,sloped] {$g_1\overline{g_2}, 0$} (q2);
    \path[->,dotted,transform canvas={xshift=0.5mm}] (q1i1) edge node[right] {$\overline{g_1}g_2, 1$} (q1);
    \draw[->,dotted,-Triangle] (q1i1) |- node[below] {$\overline{g_1}\,\overline{g_2}, 1$} (q3);

    \path[->,transform canvas={yshift=0.5mm}] (q2) edge node[above] {$\overline{r_1}, \infty$} (q2i0);
    \path[->,paredgee,bend right=55] (q2i0) edge node[above,sloped,xshift=5mm] {$\overline{g_1}g_2, 0$} (q0);
    \path[->,dotted,transform canvas={yshift=-0.5mm}] (q2i0) edge node[below] {$\overline{g_2}, 1$} (q2);
    \path[->,transform canvas={xshift=-0.5mm}] (q2) edge node[left] {$r_1, \infty$} (q2i1);
    \path[->,paredgee] (q2i1) edge node[below,xshift=10mm,sloped] {$\overline{g_1}g_2, 0$} (q1);
    \path[->,dotted,transform canvas={xshift=0.5mm}] (q2i1) edge node[right] {$g_1\overline{g_2}, 1$} (q2);
    \draw[->,dotted,-Triangle] (q2i1) |- node[below] {$\overline{g_1}\,\overline{g_2}, 1$} (q4);

%    \node[parsquare,draw,dotted] (bot) at (0,0.9,-6) {$\bot$};
%    \path[dotted,->] (bot) edge[loop right] node[right=1pt] {\tiny $\textsf{true},1$} (bot);
%    \draw[dotted,->] (q0i0) edge node[sloped,above,xshift=-8mm] {\tiny $g_1g_2,1$} (bot);
%    \draw[dotted,->] (q0i1) edge node[sloped,above,xshift=-11mm] {\tiny $g_1g_2,1$} (bot);
%    \draw[dotted,->] (q0i2) edge node[sloped,above,xshift=-2mm] {\tiny $g_1g_2,1$} (bot);
%    \draw[dotted,->] (q0i3) edge node[sloped,above,xshift=-11mm] {\tiny $g_1g_2,1$} (bot);
%
%    \draw[dotted,->] (q1i0) edge node[sloped,above,xshift=-22mm] {\tiny $g_1g_2,1$} (bot);
%    \draw[dotted,->] (q1i1) edge node[sloped,below,xshift=-22mm] {\tiny $g_1g_2,1$} (bot);
%    \draw[dotted,->] (q2i0) edge node[sloped,above,xshift=-5mm] {\tiny $g_1g_2,1$} (bot);
%    \draw[dotted,->] (q2i1) edge node[sloped,below,xshift=0mm] {\tiny $g_1g_2,1$} (bot);

\end{tikzpicture}
\end{center}
\caption{Parity game arena after two iterations of the main loop and optimal winning strategy for the controller.
For the sake of succinctness the colour $\chi(((a_0, \underline{b_1}, c_0), r_2), g_1 \overline{g_2})$ was changed from $1$ to $0$ to reduce the number of iterations.
}\label{fig:pg-ex-3}
\end{figure}

We now run the parity game solver, i.e.\ query $\owin$ for the so-far constructed arena. We first mark the boundary nodes as losing for the controller and ask $\owin$ for an optimal winning strategy for the controller. In this case, this leads to the nondeterministic strategy $\sigma$  depicted in Fig.~\ref{fig:pg-ex-2} where the thick edges in dark green belong to $\sigma$, while edges disabled by $\sigma$ are drawn as dashed lines. We will describe our instantiation of $\owin$ in Section~\ref{sec:parity-game-solver} in more detail, but intuitively $\owin$ tries to maximise the distance from the trivial losing states $\bot$ and the boundary $B$ with each colour interpreted as a distance; hence, in the intermediate states corresponding to $I\in\{\overline{r_1}\, r_2, \overline{r_1}\,\overline{r_2}, r_1\,\overline{r_2}\}$, $\owin$ chooses to let the controller play back to the initial state so that ideally never the losing state is reached and thus the distance to it is maximised.
One particular feature of our instantiation of $\owin$ is that it outputs nondeterministic strategies, i.e.\ a strategy $\sigma$ for $\penv$ is still required to respect the edge relation of the parity game but $\sigma(q)$ is only required to be some subset of the successors of $q$ in the parity game; in particular $\sigma(q)=\emptyset$ is allowed and has to be interpreted as $\pctrl$ giving up at node $q$ which is the case of the strategy shown in Fig.~\ref{fig:pg-ex-2} at the intermediate node $(q_0, r_1\,r_2)$.
Note that we do not draw $\bot$ and the corresponding edges in Fig.~\ref{fig:pg-ex-2} and Fig.~\ref*{fig:pg-ex-3} as $\owin$ will always prefer to tell the controller $\pctrl$ to give up instead of playing to $\bot$ (analogously for $\penv$ and $\top$).
More importantly, if a strategy tells a player to give up at a specific node this means that the player loses any play reaching this specific node.
For this reason, the computed strategy $\sigma$ is not winning for the controller so far, hence, we also ask $\owin$ if the environment $\penv$ might win the initial state -- now with the boundary marked as winning for the controller. As $\owin$ also fails to find a winning strategy for the environment ($\pctrl$ can easily force $\penv$ directly into the boundary), we proceed to further explore and construct the arena.

In this iteration $\oexpl$ now uses $\sigma$ and $\tau$: as neither the controller nor the environment wins the initial node so far, starting in $q_0=(a_0,\underline{b_0},c_0)$ each player can force his opponent into the boundary; hence, $\sigma$ and $\tau$ give us some information where to further explore the arena; further, from $\alpha$ we obtain scores that tell us how far from acceptance resp.\ rejection a given boundary state is w.r.t.\ to the product of the underlying automata; in our example, using $\alpha$ the oracle $\oexpl$ tells us to first explore only the two states $(a_0,\underline{b_1},c_0)$ and $(a_0,b_0,\underline{c_0})$ as in both cases only one of the two processes is waiting for being granted access.
Proceeding as before, we further extend the parity game under construction leading to the parity game shown in Fig.~\ref{fig:pg-ex-3} (which also shows the strategy obtained for $\pctrl$) and again ask if either $\pctrl$ or $\penv$ can now win the initial node $q_0$.
As the extended parity game coincides with that of the previous iteration in all but the boundary nodes, we pass the so-far computed strategies also to $\owin$ in order to re-use the information stored in them.

Fig.~\ref{fig:pg-ex-3} shows the strategy that $\owin$ now computes: the updated $\sigma$ coincides with the previous $\sigma$ on the nodes where $\pctrl$ did not give up (edges coloured in dark green); it only adds the edges coloured in light blue.
The so updated $\sigma$ now wins the initial node for the controller.
In particular, $\sigma$ keeps the nondeterminism at $(q_0,r_1\,r_2)$ where it only tells the controller to grant access to exactly one process but it does not tell $\pctrl$ which one of the processes should be preferred. This ambiguity can be used when translating the strategy into a circuit or a program to reduce the description size.
Finally note that as we mark the nodes on the boundary as losing for the respective \enquote{main} player when calling $\owin$, if the \enquote{main} player can win the so-far constructed parity game, then his strategy has to avoid the boundary, i.e.\ by construction we always find winning strategies that try to enclose all plays starting in the initial node in a \enquote{minimal} winning region. We remark that this bears some similarity to the local strategy iteration schemes by Friedmann~\cite{DBLP:journals/ijfcs/FriedmannL12}; but there the parity game is assumed to be explicitly given, and the goal is simply to speed-up strategy iteration itself; in our case the goal is to construct as little as possible from the actual parity game, while the actual choice of the oracle $\owin$ is unimportant at this point.

This brings us to the end of our walk-through of the main algorithm. In the following sections we describe in more detail how we choose to instantiate the oracles.

\subsection{DPA Construction}

\subsubsection{Formula Analysis and Decomposition}

Before constructing a DPA the formula is analysed and its syntax-tree is annotated with automata acceptance conditions based on syntactic criteria. Such a formula decomposition focussed on conjunctions has been previously used in other work such as \cite{DBLP:journals/corr/abs-1006-1408,DBLP:journals/fmsd/FiliotJR11,DBLP:journals/fmsd/Ehlers12}. However, we will also consider disjunctions and bi-implications. In Section \ref{subsec:automata} we introduced the following three sub-classes of DPAs: DWAs, DBAs, and DCAs. Accordingly we annotate the LTL formula  with \enquote{acceptance-typing} information:

\begin{definition}[Acceptance-Type Annotated LTL]
\[\begin{array}{lllllr}
\alpha  \Coloneqq \; & \phantom{\mid} \; \mathcal{B}_\wedge(\alpha, \alpha) & \mid \mathcal{C}_\wedge(\alpha, \alpha)          & \mid \mathcal{P}_\wedge(\alpha, \alpha)          & \mid \mathcal{W}_\wedge(\alpha, \alpha)          & \\
                     & \mid \mathcal{B}_\vee(\alpha, \alpha)                & \mid \mathcal{C}_\vee(\alpha, \alpha)            & \mid \mathcal{P}_\vee(\alpha, \alpha)            & \mid \mathcal{W}_\vee(\alpha, \alpha)            & \\
                     & \mid \mathcal{B}_\leftrightarrow(\alpha, \alpha)     & \mid \mathcal{C}_\leftrightarrow(\alpha, \alpha) & \mid \mathcal{P}_\leftrightarrow(\alpha, \alpha) & \mid \mathcal{W}_\leftrightarrow(\alpha, \alpha) & \\
                     & \mid \mathcal{B}(\varphi)                            & \mid \mathcal{C}(\varphi)                        & \mid \mathcal{P}(\varphi)                        & \mid \mathcal{W}(\varphi)                        & \text{ with } \varphi \in LTL
\end{array}\]
\end{definition}

We obtain an acceptance-typed LTL formula $\alpha$ from an LTL formula $\varphi$ using the following heuristic approach: First, we determine syntactically the \enquote{simplest} acceptance type, denoted $\acceptance_\varphi$, such that we can build a deterministic automaton with acceptance $\acceptance_\varphi$ for $\varphi$ efficiently. Second, we annotate $\varphi$ with this information and obtain $\alpha$ as the result of $\decompose_\varphi$. Formally:

\begin{definition}
Let $\varphi$ be a formula. Then $\acceptance$ is recursively defined as:
\[\acceptance_\varphi = \begin{cases}
    \acceptance_{\psi_1} \sqcup \acceptance_{\psi_2} & \text{if } \varphi = \psi_1 \; op \; \psi_2 \text{ with } op \, \in \{\wedge, \vee\} \\
	\mathcal{W} & \text{if } \varphi \in \ltlfum \cup \ltlgrw \\
	  & \quad \text{or if } \varphi = \psi_1 \leftrightarrow \psi_2 \text{ and }  \{\acceptance_{\psi_1}, \acceptance_{\psi_2}\} = \{\mathcal{W}\} \\
	\mathcal{B} & \text{if } \varphi \in \G(\ltlfum) \\
	\mathcal{C} & \text{if } \varphi \in \F(\ltlgrw) \\
	\mathcal{P} & \text{otherwise}
\end{cases}\]
\noindent where $\sqcup$ denotes the least upper bound relative to the partial order $\preceq$ defined by $\mathcal{W} \prec \mathcal{B}$, $\mathcal{W} \prec \mathcal{C}$, $\mathcal{B} \prec \mathcal{P}$, $\mathcal{C} \prec \mathcal{P}$, $\mathcal{B} \not\preceq \mathcal{C}$ and $\mathcal{C} \not\preceq \mathcal{B}$. The acceptance-type annotated formula $\decompose_\varphi$ is then recursively defined as:
\[\decompose_\varphi = \begin{cases}
	(\acceptance_{\varphi})_{op}(\decompose_{\psi_1}, \decompose_{\psi_2}) & \text{if } \varphi = \psi_1 \; op \; \psi_2 \text{ with } op \in \{\wedge, \vee, \leftrightarrow\} \\ & \quad \text{ and } \{\acceptance_{\psi_1}, \acceptance_{\psi_2}\} \neq \{\mathcal{P}\} \\
	\acceptance_\varphi(\varphi) & \text{otherwise} \\
\end{cases}\]
\end{definition}

In a specific implementation this decomposition and annotation might be fine-tuned to allow better translation performance, e.g. safety properties classified as  separate weak sub-formulas might be grouped for performance reasons.

Further, let $\mathcal{X}$, $\mathcal{Y}$, and $\mathcal{Z}$ be acceptance-typed LTL formulas. We make use of the following simple pattern matching notation:
\begin{itemize}
	\item $\mathcal{X} = \mathcal{B}_\land(\mathcal{X}_1, \mathcal{X}_2)$: Here $\mathcal{X}_1$ and $\mathcal{X}_2$ are fresh variables binding to the left and right subtree of $\mathcal{X}$, which is constrained to be a conjunction typed as Büchi acceptance.
	\item $\mathcal{Y} = \mathcal{Y}_{\lor}(\mathcal{C}_1, \mathcal{Y}_2)$: Here $\mathcal{C}_1$ and $\mathcal{Y}_2$ are fresh variables binding to the left and right subtree of $\mathcal{Y}$. Further, $\mathcal{C}_1$ has to be typed as co-Büchi acceptance and $\mathcal{Y}$ can be typed with any acceptance condition, but needs to be a disjunction.
	\item $\mathcal{Z} = \mathcal{W}_{\leftrightarrow}(\mathcal{W}_1, \mathcal{W}_2)$: Here $\mathcal{W}_1$ and $\mathcal{W}_2$ are fresh variables binding to the left and right subtree of $\mathcal{Z}$ and are both typed with weak acceptance. Moreover, $\mathcal{Z}$ is a bi-implication with weak acceptance.
\end{itemize}

Moreover, instead of only binary conjunctives, we use B\"uchi conjunction and co-B\"uchi disjunction of sets, i.e. we add
\begin{align*}
    \alpha \Coloneqq \; & \mathcal{B}_\land(\mathcal{B}_1, \ldots, \mathcal{B}_n) \mid \mathcal{C}_\lor(\mathcal{C}_1, \ldots, \mathcal{C}_n)
\end{align*}
for any $n \ge 2$ to the syntax. We restrict this rule to applications where all children are in the B\"uchi class for $\mathcal{B}_\land$ (resp. co-B\"uchi class for $\mathcal{C}_\lor$). After computing $\decompose_\varphi$, successive conjunctions are directly grouped together with the rule $\mathcal{B}_\land(\mathcal{B}_\land(\mathcal{B}_1, \mathcal{B}_2), \mathcal{B}_3) = \mathcal{B}_\land(\mathcal{B}_1, \mathcal{B}_\land(\mathcal{B}_2, \mathcal{B}_3)) = \mathcal{B}_\land(\mathcal{B}_1, \mathcal{B}_2, \mathcal{B}_3)$, and the respective rule for $\mathcal{C}_\lor$.

\subsubsection{Product Construction with LAR}

Given an acceptance-type annotated syntax tree $\decompose_\varphi$ for a formula $\varphi$, we now describe a recursive procedure to construct a (transition-based) DPA $\cP(\decompose_\varphi)$ with $\lang(\cP(\decompose_\varphi)) = \lang(\varphi)$. Observe that not all patterns that are syntactically possible are covered, but all patterns generated by $\decompose_\varphi$.

\paragraph{Base Case.}

In the case $\decompose_\varphi \in \{\mathcal{W}(\varphi), \mathcal{B}(\varphi), \mathcal{C}(\varphi), \mathcal{P}(\varphi)\}$ we use one of the direct automata constructions described in Section \ref{subsec:ltl-fragments}.

\paragraph{Conjunction.}

Now consider the case for the conjunction $\decompose_{\varphi} = \mathcal{X}_\land(\mathcal{T}_{\psi_1}, \mathcal{T}_{\psi_2})$, or $\mathcal{T}_{\varphi} = \mathcal{B}_\land(\mathcal{T}_{\psi_1}, \ldots, \mathcal{T}_{\psi_n})$.
We start the construction by recursively constructing DPAs $\cP(\decompose_{\psi_i}) = (Q_i, \delta_i, q^0_i, \chi_i, d_i, \parity_i)$ for each child $\decompose_{\psi_i}$.
Then we apply a case distinction based on $\decompose_{\varphi}$ and each $\decompose_{\psi_i}$.

\begin{itemize}
	\item

In the cases where we have $\mathcal{X}_\land(\mathcal{W}_1, \mathcal{X}_2)$ (and symmetrically $\mathcal{X}_\land(\mathcal{X}_1, \mathcal{W}_2)$),
we can use a simple product construction.
Then define the DPA
\begin{align*}
    \cP(\mathcal{X}_\land(\mathcal{W}_1, \mathcal{X}_2)) &\coloneqq (Q_1 \times Q_2, \delta, (q^0_1, q^0_2), \chi, d_2, \parity_2)
\end{align*}
with:
\begin{align*}
    \delta(q, a) &\coloneqq (\delta_1(q_1, a), \delta_2(q_2, a)) &
    \chi(q, a) &\coloneqq
    \begin{cases}
        \chi_2(q_2, a) & \text{if $\chi_1(q_1, a) = \parity_1$} \\
        \overline{\parity_2} & \text{if $\chi_1(q_1, a) \neq \parity_1$}
    \end{cases}
\end{align*}

\item In the cases where we have $\mathcal{X}_\land(\mathcal{C}_1, \mathcal{X}_2)$ (and symmetrically $\mathcal{X}_\land(\mathcal{X}_1, \mathcal{C}_2)$),
we can also use a product construction, with possibly one extra colour.
W.l.o.g. assume $\parity_2 = \odd$. This can be achieved by switching the parity of $A(\mathcal{X}_2)$ if necessary.
Then define the DPA
\begin{align*}
    \cP(\mathcal{X}_\land(\mathcal{C}_1, \mathcal{X}_2)) &\coloneqq (Q_1 \times Q_2, \delta, (q^0_1, q^0_2), \chi, d_2, 1)
\end{align*}
with:
\begin{align*}
    \delta(q, a) &\coloneqq (\delta_1(q_1, a), \delta_2(q_2, a)) &
    \chi(q, a) &\coloneqq
    \begin{cases}
        0 & \text{if $\chi_1(q_1, a) = 0$} \\
        \chi_2(q_2, a) & \text{if $\chi_1(q_1, a) = 1$}
    \end{cases}
\end{align*}

\item Next, we consider the case $\mathcal{B}_\land(\mathcal{B}_1, \ldots, \mathcal{B}_n)$ with two or more B\"uchi children, and only B\"uchi children.
Here, on top of a product construction, we need an additional \emph{round-robin-counter} to track of successive
satisfaction of the B\"uchi acceptance of the children.
We define the DBA
\begin{align*}
    \cP(\mathcal{B}_\land(\mathcal{B}_1, \ldots, \mathcal{B}_n)) &\coloneqq (Q \times \{ 0, 1, \ldots, n-1 \}, \delta, (q^0, 0), \chi, 1, \even)
\end{align*}
with $Q \coloneqq \left( Q_1 \times \ldots \times Q_n \right)$, $q^0 \coloneqq (q^0_1, \ldots, q^0_n)$ and
\begin{align*}
    \delta((q, r), a) &\coloneqq ((\delta_1(q_1, a), \ldots, \delta_n(q_n, a)), r' \bmod n) \\
    \chi((q, r), a) &\coloneqq
    \begin{cases}
        0 & \text{if $r' = n$} \\
        1 & \text{if $r' < n$}
    \end{cases}
\end{align*}
where $r' \coloneqq \max\{ s \in \{r,r+1,\ldots,n\} \mid \forall r < j \le s : \chi_j(q_j, a) = \even \}$.

\item
Last, we consider the case $\mathcal{P}_\land(\mathcal{B}_1, \mathcal{P}_2)$ (and symmetrically $\mathcal{P}_\land(\mathcal{P}_1, \mathcal{B}_2)$).
W.l.o.g. we may assume $\parity_2 = \odd$ by switching parity if necessary.
Here, we need additional memory to remember the \emph{minimal colour} of $\mathcal{P}_2$ between acceptances of $\mathcal{B}_1$.
We define the DPA
\begin{align*}
    \cP(\mathcal{P}_\land(\mathcal{B}_1, \mathcal{P}_2)) &\coloneqq (Q_1 \times Q_2 \times \{0, 1, \ldots, d_2\}, \delta, ((q^0_1, q^0_2), d_2), \chi, d, 1)
\end{align*}
with $d \coloneqq \min\{d \in \{d_2, d_2 + 1\} \mid d \equiv_2 \even \}$ and
\begin{align*}
    \delta((q, c), a) &\coloneqq
    \begin{cases}
        ((\delta_1(q_1, a), \delta_2(q_2, a)), d_2) & \text{if $\chi_1(q_1, a) = 0$} \\
        ((\delta_1(q_1, a), \delta_2(q_2, a)), c') & \text{otherwise}
    \end{cases} \\
    \chi((q, c), a) &\coloneqq
    \begin{cases}
        c' & \text{if $\chi_1(q_1, a) = 0$} \\
        d  & \text{otherwise}
    \end{cases}
\end{align*}
where $c' \coloneqq \min\left( c, \chi_2(q_2, a) \right)$.

\end{itemize}

Note that if some child in a conjunction reaches a non-accepting sink, then we also
know that the conjunction can never accept again, and we can simplify the product state.
A similar argument holds if all children reach a accepting sink.
Formally, we replace $\delta$ by $\delta'$ and $\chi$ by $\chi'$ defined by:
\begin{align*}
    \delta'(q, a) &\coloneqq
    \begin{cases}
        q & \text{if $q \in \{\bot,\top\}$} \\
        \top & \text{if for all $i$ we have $\delta_i(q_i, a) = \top$} \\
        \bot & \text{if for some $i$ we have $\delta_i(q_i, a) = \bot$} \\
        \delta(q, a) & \text{otherwise}
    \end{cases} \\
    \chi'(q, a) &\coloneqq
    \begin{cases}
        \parity & \text{if $\delta'(q,a) = \top$} \\
        \overline{\parity} & \text{if $\delta'(q,a) = \bot$} \\
        \chi(q, a) & \text{otherwise}
    \end{cases}
\end{align*}

\paragraph{Disjunction.}  The construction of the DPA $\cP(\mathcal{X}_\lor(\mathcal{X}_1, \mathcal{X}_2))$
or $\cP(\mathcal{C}_\lor(\mathcal{C}_1, \ldots, \mathcal{C}_n))$
for the disjunctive $\mathcal{X}_\lor$ is dual to the conjunction case.

\paragraph{Bi-implication.} Finally, we consider the bi-implication $\mathcal{X}_\leftrightarrow(\mathcal{X}_1, \mathcal{X}_2)$.
This can be expressed through
$\mathcal{X}_\land$ and $\mathcal{X}_\lor$ by the logical equivalence $\varphi \leftrightarrow \psi \equiv (\varphi \land \psi) \lor (\lnot \varphi \land \lnot \psi)$.
However, this construction would increase the state space and number of colours in some cases, since four automata ($\varphi$, $\neg \varphi$, $\psi$, $\neg \psi$) instead of two ($\varphi$, $\psi$) need to be constructed.
Therefore we have a special construction for the DPA $\cP(\mathcal{X}_\leftrightarrow(\mathcal{X}_1, \mathcal{X}_2))$.
As before, we start by constructing the DPAs for the children $\mathcal{X}_1,\mathcal{X}_2$.
Let $\cP(\mathcal{X}_i) = (Q_i, \delta_i, q^0_i, \chi_i, d_i, \parity_i)$ for $i \in \{1,2\}$.

\begin{itemize}
\item
In the case $\mathcal{W}_\leftrightarrow(\mathcal{W}_1, \mathcal{W}_2)$,
we can apply a simple product construction.
W.l.o.g. assume $\parity_1 = \parity_2$ by switching parity if necessary.
We define the DWA
\begin{align*}
    \cP(\mathcal{W}_\leftrightarrow(\mathcal{W}_1, \mathcal{W}_2)) &\coloneqq (Q_1 \times Q_2, \delta, (q^0_1, q^0_2), \chi, 1, \even)
\end{align*}
with:
\begin{align*}
    \delta(q, a) &\coloneqq (\delta_1(q_1, a), \delta_2(q_2, a)) &
    \chi(q, a) &\coloneqq \left(\chi_1(q_1, a) + \chi_2(q_2, a)\right) \bmod 2
\end{align*}
\item
    Now we consider the general case for $\mathcal{P}_\leftrightarrow(\mathcal{X}_1, \mathcal{X}_2)$ (note that $\mathcal{B}_\leftrightarrow$ and $\mathcal{C}_\leftrightarrow$ never occur).
W.l.o.g. assume that $\mathcal{X}_1 \neq \mathcal{P}$ and thus $d_1 = 1$.
We consider $\mathcal{X}_2$ to be of class $\mathcal{P}$ and need to store its \emph{minimal colour} between acceptances of $\mathcal{X}_1$, as for conjunction.
However, whenever $\mathcal{X}_1$ does not accept, we emit the colour of $A_2$, shifted by one.
If $\mathcal{X}_1 = \mathcal{W}$, we can actually omit the memory to store the colour of $\mathcal{X}_2$.
We define the DPA
\begin{align*}
    \cP(\mathcal{P}_\leftrightarrow(\mathcal{X}_1, \mathcal{X}_2)) &\coloneqq (\left( Q_1 \times Q_2 \right) \times \{ 0, 1, \ldots, d_2 \}, \delta, ((q^0_1, q^0_2), d_2), \chi, d, \parity)
\end{align*}
with $d \coloneqq d_2 + 1$, $\parity \coloneqq \left(\parity_1 + \parity_2\right) \bmod 2$ and
\begin{align*}
    \delta((q, c), a) &\coloneqq
    \begin{cases}
        ((\delta_1(q_1, a), \delta_2(q_2, a)), d_2) & \text{if $\chi_1(q_1, a) = \parity_1$ or $\mathcal{X}_1 = \mathcal{W}$} \\
%        ((\delta_1(q_1, a), \delta_2(q_2, a)), \min \left( c, \chi_2(q_2, a) \right)) & \text{if $\chi_1(q_1, a) \neq \parity_1$}
        ((\delta_1(q_1, a), \delta_2(q_2, a)), c') & \text{if $\chi_1(q_1, a) \neq \parity_1$}
    \end{cases} \\
    \chi(q, a) &\coloneqq
    \begin{cases}
        c' & \text{if $\chi_1(q_1, a) = \parity_1$} \\
        \chi_2(q_2, a) + 1 & \text{if $\chi_1(q_1, a) \neq \parity_1$}
    \end{cases}
\end{align*}
where $c' \coloneqq \min\left( c, \chi_2(q_2, a) \right)$.

\item We note that the construction of the previous case can be generalized to the product of two arbitrary DPAs by remembering for each colour $c$ of $\cP(\mathcal{X}_1)$
the minimal colour of $\cP(\mathcal{X}_2)$ between minimal occurrences of $c$ in $\cP(\mathcal{X}_1)$.
\end{itemize}

As with conjunction, we apply the simplification to $\cP(\mathcal{X}_\leftrightarrow(\mathcal{X}_1,\mathcal{X}_2))$,
that if both children reach a state in $\{\bot,\top\}$, also the product state is either
$\bot$ or $\top$.
Replace $\delta$ by $\delta'$ and $\chi$ by $\chi'$ defined by:
\begin{align*}
    \delta'(q, a) &\coloneqq
    \begin{cases}
        q & \text{if $q \in \{\bot,\top\}$} \\
        \top & \text{if $\delta_1(q_1, a),\delta_2(q_2,a) \in \{\bot, \top\}$ and $\delta_1(q_1,a) = \delta_2(q_2,a)$} \\
        \bot & \text{if $\delta_1(q_1, a),\delta_2(q_2,a) \in \{\bot, \top\}$ and $\delta_1(q_1,a) \neq \delta_2(q_2,a)$} \\
        \delta(q, a) & \text{otherwise}
    \end{cases} \\
    \chi'(q, a) &\coloneqq
    \begin{cases}
        \parity & \text{if $\delta'(q,a) = \top$} \\
        \overline{\parity} & \text{if $\delta'(q,a) = \bot$} \\
        \chi(q, a) & \text{otherwise}
    \end{cases}
\end{align*}

Let $\varphi$ be a formula and let $\alpha = \decompose_\varphi$ be the acceptance-type annotated formula. 
We then implement the oracles $\oparity$, $\oinit$, and $\osucc$ by the DPA $\cP(\alpha) = (Q, \delta, q^0, \chi, d, \parity)$ in the following way:
$\oparity(\alpha) \coloneqq \parity$,
$\oinit(\alpha) \coloneqq q^0$, and
$\osucc(\alpha, q) \coloneqq \{ (\{i\}, \{o\}, \chi(q, i \cup o), \delta(q, i \cup o) ) \mid i \in \alin, o \in \alout \}$.

\subsubsection{Implementation Details}

\paragraph{On-the-fly construction.}
For this construction, the functions $\delta$ and $\chi$ only need to query the local state, and never need a global state.
They might call the functions $\delta$ and $\chi$ for their children, but those also only depend on the local state and their respective children.
Therefore it is possible to implement the construction of the DPA $\cP(\alpha)$ for a decomposition $\alpha$ \emph{on-the-fly}: Starting with the initial state,
successor states are only generated when necessary, and the DPA is not fully constructed until all states have been queried.
This holds both for the root automaton and for any child automata constructed by the decomposition.

\paragraph{Memoization.} When querying the root DPA for successors, the successors of the same state in a child DPA may be needed several times.
Instead of recomputing them each time, the successors of the state are \emph{cached} or \emph{memoized} for direct access.

\paragraph{Formula Isomorphism.} Building up on the memoization feature the construction only constructs one automaton for a pair of formulas isomorphic under renaming atomic propositions und remaps letters in the query stage, effectively reducing the automata states needed to be constructed for parametric formulas, where the same pattern is repeated with different atomic propositions.

\paragraph{Symbolic Construction and Representation of Transition Relations.} The description of the oracle interface only specifies that successor function represented by a set needs to be returned. An implementation can choose to represent such a transition relation explicitly (e.g. by a list of length $2^n$) or symbolically (e.g. by an MTBDD). In fact \toolname{} 19.07 relies on a symbolic construction and representation of the transition relation of the arena and the automata. This then allows efficient (symbolic) grouping of inputs and outputs into equivalence classes by $\osucc$.

\subsection{Parity Game Solver}\label{sec:parity-game-solver}
We instantiate $\owin$ with a variant of the strategy iteration algorithm in~\cite{DBLP:journals/corr/abs-0806-2923}. 

We first give a brief description of what strategy iteration is and why we deem it particularly useful when combined with a demand-driven construction of the arena. We then exemplify these ideas in a bit more detail using our preceding example from Sec.~3.

In brief, strategy iteration consists of improving the current strategy (resp.\ controller) by iterating the two steps: (i) compute the \enquote{worst case} the environment can inflict on the system w.r.t.\ the current strategy; (ii) state-wise redefine the current strategy by selecting any successor(s) which lead to a better \enquote{worst case}. The iteration stops as soon as the strategy cannot be improved anymore i.e.\ if the currently selected successors led to the best \enquote{worst case}. Strategy iteration nicely combines with our approach of constructing the actual arena in a demand driven way: we only construct the arena to that extent that allows the controller to stay within the constructed subarena; all nodes resp.\ edges which are outside of this subarena are simply flagged to be lost to the controller, when checking for realisability, resp.\ the system, when checking for unrealisability; in addition, if we need to further explore the arena as we could neither prove realisability nor unrealisability using the subarena constructed so far, we can re-use the already computed strategies as initial strategies for computing the optimal strategies for the extended subarena.

We exemplify these ideas now.
Let $\cA=(\vctrl,\venv, E, \chi, B)$ be a parity game arena with nodes $V=\vctrl\cup \venv$ split between the two players $\pctrl$ and $\penv$, edges $E$, edge colouring $\chi\colon E\to \N_0$, and boundary nodes $B\subseteq V$.
We assume that $\cA$ includes two special nodes $\bot$ and $\top$ where $\bot$ is always won by the environment $\penv$ and $\top$ is always won by the controller $\pctrl$.
We further require that all nodes in $V\setminus B$ have at least one successor w.r.t.\ $E$.\footnote{Actually, the edges are also labeled by corresponding inputs and outputs coming from the environment and the controller, respectively.}
To simplify notation, we forget about the inputs and outputs that also label the edges (as shown e.g.\ in Fig.~\ref{fig:pg-ex-1}) s.t.\ we can simply write $vE$ for the set of successors of the node $v$.

Besides $\cA$, the parity game solver takes as additional input the \enquote{main player} $\plr\in \{\pctrl,\penv\}$, 
the parity $\parity\in\{0,1\}$ with which $\plr$ wins a play, 
a node $q\in V$ of the arena whose winner we want to determine,
and an initial strategy $\kappa$ for the main player $\plr$. We will write $\overline{\plr}$ for the opponent of $\plr$ s.t.\ $\{\plr,\opp\}=\{\pctrl,\penv\}$ with the parity of the opponent $\opp$ being $\overline{\parity}$ accordingly.
All nodes in $B$ are considered to be losing for the main player $P$, i.e.\ $\overline{\plr}$ can win be forcing $\plr$ into $B$. 

In order to solve the parity game, we reformulate the winning condition into a sup-inf-distance problem: 
To this end, we first introduce an (implicit) auxiliary node $\bullet$ to which only the main player $\plr$ can move to in order to {\em give up}. 
This node is in addition to the two nodes $\top$ and $\bot$ and only serves to simplify presentation.
All edges leading to $\bullet$ are defined to be coloured by $\infty$ (\enquote{don't care}).
We denote such a modification of an arena $\cA$ by $\cA^\bullet$.

We interpret the edge colours $\chi(v,w)$ as weights $\gamma(v,w)$ that measure how close $P$ comes to winning resp.\ losing when taking the corresponding edge.
Let $C\subseteq \N_0$ be the set of all colours occurring in $\cA$ except for $\infty$. (Recall that we used the colour \enquote{$\infty$} for edges that are unimportant w.r.t.\ the winning condition.) 
A colour $c\in C$ is identified with the multiset $\{c\}$ which we represent by its characteristic function w.r.t.\ $\N_0^C$. The colour $\infty$ is identified with the empty multiset resp.\ its characteristic function.
Addition on $\N_0^C$ is defined point-wise as usual s.t.\ it coincides with the union of multisets.
The weight of a finite play is then simply the sum of the weights of the edges traversed by it, i.e.\ the multiset of the colours of the edges of a play.

We order $\N_0^C$ from the point of view of $\plr$ by means of the following order relation $\pord$: Given two distinct functions $g,g'\in \N_0^C$, let $c:=\min \{ c' \in C \mid g(c') \neq g'(c')\}$ be the least colour in which the two differ;
if $c$ has parity $\parity$, we set $g \pord g'$ if and only if $g(c) < g'(c)$; else if $c$ has parity $\overline{\parity}$, we set $g \pord g'$ if and only if $g(c)> g'(c)$.
For instance, taking a look at the arena of Fig.~\ref{fig:pg-ex-1}, the colours $0$ and $1$ are mapped on the functions (represented as tuples) $(1,0)$ and $(0,1)$, respectively. From the perspective of the controller $\pctrl$ and its winning parity $0$, the weight $(1,0)$ (representing the colour $0$) is more attractive than the weight $(0,1)$ (representing the colour $1$), i.e.\ $(1,0) \pord (0,1)$.
%Consider e.g.\ two finite plays both resulting from finite cycles. If in both cycles the least colour has parity $\parity$, then $\plr$ will prefer the cycle whose least colour is smaller, or, if both are equal, the cycle in the which the least colour occurs more often.
%Symmetrically, if the partiy of both cycles, he will prefer the cycle with larger least colour, or, the least colour be the same in both cycles, the one which visits the least colour less often.
%Finally, should the parity of the least colour differ in both cycles, $\plr$ will by the cylce in which the least colour has parity $\parity$.
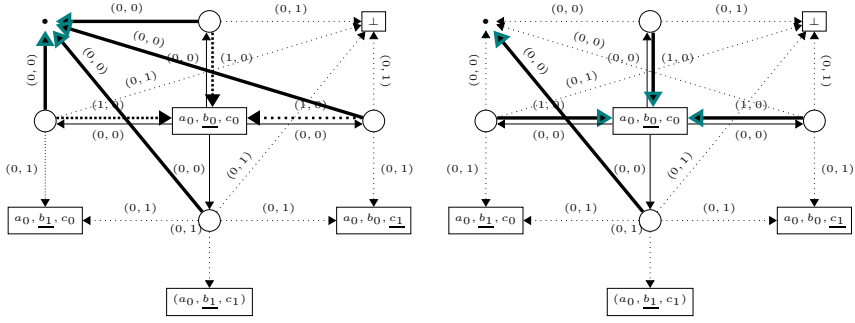
\begin{figure}[!ht]
\begin{center}
\scalebox{0.8}{
\begin{tikzpicture}[scale=1.5,font={\tiny}]
    \node[parsquare,draw] (q0) at (0,2) {$a_0,\underline{b_0},c_0$};
    \node[parcircle,draw] (q0i0) at (0,3.1) {};
    \node[parcircle,draw] (q0i1) at (-1.8,2) {};
    \node[parcircle,draw] (q0i2) at (1.8,2) {};
    \node[parcircle,draw] (q0i3) at (0,0.9) {};

    \node[parsquare,draw] (q1) at (-1.8,0.9) {$a_0,\underline{b_1},c_0$};

    \node[parsquare,draw] (q2) at (1.8,0.9) {$a_0,b_0,\underline{c_1}$};

    \node[parsquare,draw] (q3) at (0,0.0) {$(a_0,\underline{b_1},c_1)$};

    \path[->,transform canvas={xshift=-0.5mm}] (q0) edge node[left,yshift=2mm] {$(0,0)$} (q0i0);
    \path[very thick,->,densely  dotted,transform canvas={xshift=0.5mm}] (q0i0) edge node[right,yshift=2mm] {$(1,0)$} (q0);
    \path[very thick,->,densely  dotted,transform canvas={yshift=0.5mm}] (q0i1) edge node[above,xshift=-1mm] {$(1,0)$} (q0);
    \path[->,transform canvas={yshift=-0.5mm}] (q0) edge node[below,xshift=-1mm] {$(0,0)$} (q0i1);
    \path[->,densely  dotted] (q0i1) edge node[left] {$(0,1)$} (q1);    
    \path[->,transform canvas={yshift=-0.5mm}] (q0) edge node[below,xshift=1mm] {$(0,0)$} (q0i2);
    \path[very thick,->,dotted,transform canvas={yshift=0.5mm}] (q0i2) edge node[above,xshift=1mm] {$(1,0)$} (q0);    
    \path[->,dotted] (q0i2) edge node[right] {$(0,1)$} (q2);
    \path[->] (q0) edge node[left,xshift=0.5mm,yshift=0.5mm] {$(0,0)$} (q0i3);
    \path[->,dotted] (q0i3) edge node[above] {$(0,1)$} (q1);
    \path[->,dotted] (q0i3) edge node[above] {$(0,1)$} (q2);
    \path[->,dotted] (q0i3) edge node[left,yshift=5mm] {$(0,1)$} (q3);
    
    \node (dot) at (-1.8,3.1) {$\bullet$};
%    \path[dotted,->] (bot) edge[loop right] node[right=1pt] {\tiny $\textsf{true},1$} (dot);
    \draw[->, paredge] (q0i0) edge node[sloped,above,xshift=0mm] {\tiny $(0,0)$} (dot);
    \draw[->, paredge] (q0i1) edge node[sloped,above,xshift=0mm] {\tiny $(0,0)$} (dot);
    \draw[->, paredge] (q0i2) edge node[sloped,above,xshift=-11mm] {\tiny $(0,0)$} (dot);
    \draw[->, paredge] (q0i3) edge node[sloped,above,xshift=-11mm] {\tiny $(0,0)$} (dot);
    
    \node[parsquare] (bot) at (1.8,3.1) {$\bot$};
%    \path[dotted,->] (bot) edge[loop right] node[right=1pt] {\tiny $\textsf{true},1$} (bot);
    \draw[dotted,->] (q0i0) edge node[sloped,above,xshift=0mm] {\tiny $(0,1)$} (bot);
    \draw[dotted,->] (q0i1) edge node[sloped,above,xshift=-11mm] {\tiny $(0,1)$} (bot);
    \draw[dotted,->] (q0i2) edge node[sloped,above,xshift=0mm] {\tiny $(0,1)$} (bot);
    \draw[dotted,->] (q0i3) edge node[sloped,above,xshift=-11mm] {\tiny $(0,1)$} (bot);
\end{tikzpicture}
}
\scalebox{0.8}{
\begin{tikzpicture}[scale=1.5,font={\tiny}]
    \node[parsquare,draw] (q0) at (0,2) {$a_0,\underline{b_0},c_0$};
    \node[parcircle,draw] (q0i0) at (0,3.1) {};
    \node[parcircle,draw] (q0i1) at (-1.8,2) {};
    \node[parcircle,draw] (q0i2) at (1.8,2) {};
    \node[parcircle,draw] (q0i3) at (0,0.9) {};

    \node[parsquare,draw] (q1) at (-1.8,0.9) {$a_0,\underline{b_1},c_0$};

    \node[parsquare,draw] (q2) at (1.8,0.9) {$a_0,b_0,\underline{c_1}$};

    \node[parsquare,draw] (q3) at (0,0.0) {$(a_0,\underline{b_1},c_1)$};

    \path[->,transform canvas={xshift=-0.5mm}] (q0) edge node[left,yshift=2mm] {$(0,0)$} (q0i0);
    \path[->,paredge,transform canvas={xshift=0.5mm}] (q0i0) edge node[right,yshift=2mm] {$(1,0)$} (q0);
    \path[->,paredge,transform canvas={yshift=0.5mm}] (q0i1) edge node[above,xshift=-1mm] {$(1,0)$} (q0);
    \path[->,transform canvas={yshift=-0.5mm}] (q0) edge node[below,xshift=-1mm] {$(0,0)$} (q0i1);
    \path[->,dotted] (q0i1) edge node[left] {$(0,1)$} (q1);    
    \path[->,transform canvas={yshift=-0.5mm}] (q0) edge node[below,xshift=1mm] {$(0,0)$} (q0i2);
    \path[->,paredge,transform canvas={yshift=0.5mm}] (q0i2) edge node[above,xshift=1mm] {$(1,0)$} (q0);    
    \path[->,dotted] (q0i2) edge node[right] {$(0,1)$} (q2);
    \path[->] (q0) edge node[left,xshift=0.5mm,yshift=0.5mm] {$(0,0)$} (q0i3);
    \path[->,dotted] (q0i3) edge node[above] {$(0,1)$} (q1);
    \path[->,dotted] (q0i3) edge node[above] {$(0,1)$} (q2);
    \path[->,dotted] (q0i3) edge node[left,yshift=5mm] {$(0,1)$} (q3);
    
    \node (dot) at (-1.8,3.1) {$\bullet$};
%    \path[dotted,->] (bot) edge[loop right] node[right=1pt] {\tiny $\textsf{true},1$} (dot);
    \draw[dotted,->] (q0i0) edge node[sloped,above,xshift=0mm] {\tiny $(0,0)$} (dot);
    \draw[dotted,->] (q0i1) edge node[sloped,above,xshift=0mm] {\tiny $(0,0)$} (dot);
    \draw[dotted,->] (q0i2) edge node[sloped,above,xshift=-11mm] {\tiny $(0,0)$} (dot);
    \draw[->, paredge] (q0i3) edge node[sloped,above,xshift=-11mm] {\tiny $(0,0)$} (dot);
    
    \node[parsquare] (bot) at (1.8,3.1) {$\bot$};
%    \path[dotted,->] (bot) edge[loop right] node[right=1pt] {\tiny $\textsf{true},1$} (bot);
    \draw[dotted,->] (q0i0) edge node[sloped,above,xshift=0mm] {\tiny $(0,1)$} (bot);
    \draw[dotted,->] (q0i1) edge node[sloped,above,xshift=-11mm] {\tiny $(0,1)$} (bot);
    \draw[dotted,->] (q0i2) edge node[sloped,above,xshift=0mm] {\tiny $(0,1)$} (bot);
    \draw[dotted,->] (q0i3) edge node[sloped,above,xshift=-11mm] {\tiny $(0,1)$} (bot);
\end{tikzpicture}
}
\end{center}
\caption{\footnotesize
The arena of Fig.~\protect\ref{fig:pg-ex-1} extended by the auxiliary node $\bullet$.
The colours $0$, $1$ and $\infty$ have been transformed into the respective edge weights $(1,0)$, $(0,1)$ and $(0,0)$ in $\N_0^C$.
We assume that we have to decide whether the controller $\plr=\protect\pctrl$ can win in the initial node $(a_0,\underline{b_0},c_0)$.
Here, the parity of $\plr$ is $\parity=0$ so the colour $0$ resp.\ the tuple $(1,0)$ is positive from the point of view of $\protect\pctrl$, while the colour $1$ resp.\ the tuple $(0,1)$ is negative.
When not given an initial strategy, we start with the strategy that tells $\plr$ to give up at every node controlled by $\plr$, thereby preventing the existence of negative cycles in any case, as shown in the figure on the left.
This leads to the sup-inf-distance to be $(0,0)$ at every node s.t.\ playing back to the initial node is an improvement as it closes in each case a cycle of positive weight. Choosing all these improvements (thick dotted edges) yields the strategy shown in the figure on the right; after removing the auxiliary node $\bullet$, this yields the strategy shown already in Fig.~\protect\ref{fig:pg-ex-2}.
}
\label{fig:pg-solve-1}
\end{figure}

To make the interpretation of the functions in $\N_0^C$ as weights more intuitive, let us remark that one can recover the order $\pord$ by reading the function $g\in\N_0^C$ as a numeral w.r.t.\ the alternating basis $-b$, i.e.\ $g$ is interpreted as the integer $\sum_{c\in C} g(c)\cdot (-1)^{p+c} \cdot b^{-c+\max C}$ where $b$ is any sufficiently large positive integer, e.g.\ $b=\abs{V}$. This ensures that in every simple cycle in $\cA$ we have that the cycle is won by $\plr$ if and only if the total weight of the cycle is positive, i.e.\ staying forever in the cycle leads to gaining infinite distance. For instance in the arena of Fig.~\ref{fig:pg-solve-1} we could choose $b=2$ s.t.\ $(1,0)$ is mapped onto $2$, while $(0,1)$ is mapped on $-1$.

$\plr$'s goal thus becomes to maximise the distance to losing, i.e.\ the minimal total weight accumulated along a play -- thus $P$ will only accept infinite plays if these yield an infinite distance to losing, otherwise $P$ will use the option to play to $\bullet$ (or into the boundary) in order to terminate a play and bound the distance to losing to a finite value.

\begin{figure}[!ht]
\begin{center}
\scalebox{0.75}{
\begin{tikzpicture}[scale=1.5,font={\tiny}]
    \node[parsquare,draw] (q0) at (0,2) {$a_0,\underline{b_0},c_0$};
    \node[parcircle,draw] (q0i0) at (0,3.1) {};
    \node[parcircle,draw] (q0i1) at (-1.8,2) {};
    \node[parcircle,draw] (q0i2) at (1.8,2) {};
    \node[parcircle,draw] (q0i3) at (0,0.9) {};

    \node[parsquare,draw] (q1) at (-1.8,0.9) {$a_0,\underline{b_1},c_0$};
    \node[parcircle,draw] (q1i0) at (-3.4,0.9) {};
    \node[parcircle,draw] (q1i1) at (-1.8,0) {};

    \node[parsquare,draw] (q2) at (1.8,0.9) {$a_0,b_0,\underline{c_1}$};
    \node[parcircle,draw] (q2i0) at (3.4,0.9) {};
    \node[parcircle,draw] (q2i1) at (1.8,0) {};

    \node[parsquare,draw] (q3) at (0,-0.3) {$(a_0,\underline{b_1},c_1)$};
    
     \node[parsquare,draw] (q4) at (2.9,-0.2) {$(a_0,b_1,\underline{c_1})$};

%   \path[->,transform canvas={xshift=-0.5mm}] (q0) edge node[left,yshift=2mm] {$(0,0)$} (q0i0);
   \path[->,transform canvas={xshift=-0.5mm}] (q0) edge  (q0i0);
   \path[->,paredge,transform canvas={xshift=0.5mm}] (q0i0) edge node[right,yshift=2mm] {$(1,0)$} (q0);
   \path[->,paredge,transform canvas={yshift=0.5mm}] (q0i1) edge node[above,xshift=-1mm] {$(1,0)$} (q0);
%   \path[->,transform canvas={yshift=-0.5mm}] (q0) edge node[below,xshift=-1mm] {$(0,0)$} (q0i1);
   \path[->,transform canvas={yshift=-0.5mm}] (q0) edge  (q0i1);
   \path[->,dotted] (q0i1) edge node[left] {$(0,1)$} (q1);    
%   \path[->,transform canvas={yshift=-0.5mm}] (q0) edge node[below,xshift=1mm] {$(0,0)$} (q0i2);
   \path[->,transform canvas={yshift=-0.5mm}] (q0) edge (q0i2);
   \path[->,paredge,transform canvas={yshift=0.5mm}] (q0i2) edge node[above,xshift=1mm] {$(1,0)$} (q0);    
   \path[->,dotted] (q0i2) edge node[left] {$(0,1)$} (q2);
%   \path[->] (q0) edge node[left,xshift=0.5mm,yshift=0.5mm] {$(0,0)$} (q0i3);
   \path[->] (q0) edge  (q0i3);
   \path[->,dotted] (q0i3) edge node[above] {$(0,1)$} (q1);
   \path[->,dotted] (q0i3) edge node[above] {$(0,1)$} (q2);
   \path[->,dotted] (q0i3) edge node[left,yshift=5mm] {$(0,1)$} (q3);

%   \path[->,transform canvas={yshift=0.5mm}] (q1) edge node[above] {$(0,0)$} (q1i0);
   \path[->,transform canvas={yshift=0.5mm}] (q1) edge  (q1i0);
   \path[->,very thick,densely dotted,bend left=55] (q1i0) edge node[above,sloped,xshift=-3mm] {$(1,0)$} (q0);
   \path[->,dotted,transform canvas={yshift=-0.5mm}] (q1i0) edge node[below] {$(0,1)$} (q1);
%   \path[->,transform canvas={xshift=-0.5mm}] (q1) edge node[left] {$(0,0)$} (q1i1);
   \path[->,transform canvas={xshift=-0.5mm}] (q1) edge (q1i1);
   \path[->,very thick,densely dotted] (q1i1) edge node[below,xshift=-4mm,sloped] {$(1,0)$} (q2);
   \path[->,dotted,transform canvas={xshift=0.5mm}] (q1i1) edge node[right] {$(0,1)$} (q1);
   \draw[->,dotted,-Triangle] (q1i1) |- node[below] {$(0,1)$} ([xshift=-2mm]q3.west);

%   \path[->,transform canvas={yshift=0.5mm}] (q2) edge node[above] {$(0,0)$} (q2i0);
   \path[->,transform canvas={yshift=0.5mm}] (q2) edge (q2i0);
   \path[->,very thick,densely dotted,bend right=55] (q2i0) edge node[above,sloped,xshift=3mm] {$(1,0)$} (q0);
   \path[->,dotted,transform canvas={yshift=-0.5mm}] (q2i0) edge node[below] {$(0,1)$} (q2);
%   \path[->,transform canvas={xshift=-0.5mm}] (q2) edge node[left] {$(0,0)$} (q2i1);
   \path[->,transform canvas={xshift=-0.5mm}] (q2) edge (q2i1);
   \path[->,very thick,densely dotted] (q2i1) edge node[below,xshift=4mm,sloped] {$(1,0)$} (q1);
   \path[->,dotted,transform canvas={xshift=0.5mm}] (q2i1) edge node[right] {$(1,0)$} (q2);
%    \draw[->,dotted,-Triangle] (q2i1) |- ([xshift=2mm]q3.east);
   	\draw[->,dotted,-Triangle] (q2i1) |- node[below] {$(0,1)$} (q4);
    
    \node (dot) at (-1.8,3.1) {$\bullet$};
%    \path[dotted,->] (bot) edge[loop right] node[right=1pt] {\tiny $\textsf{true},1$} (dot);
    \draw[dotted,->] (q0i0) edge  (dot);
    \draw[dotted,->] (q0i1) edge (dot);
    \draw[dotted,->] (q0i2) edge (dot);
    \draw[->, paredge] (q0i3) edge (dot);
    \draw[->, paredge, bend left=45] (q1i0) edge (dot);
    \draw[->, paredge,bend left=60] (q1i1) edge (dot);
    \draw[->, paredge] (q2i0) edge (dot);
    \draw[->, paredge] (q2i1) edge (dot);

\end{tikzpicture}}
\scalebox{0.75}{
\begin{tikzpicture}[scale=1.5,font={\tiny}]
    \node[parsquare,draw] (q0) at (0,2) {$a_0,\underline{b_0},c_0$};
    \node[parcircle,draw] (q0i0) at (0,3.1) {};
    \node[parcircle,draw] (q0i1) at (-1.8,2) {};
    \node[parcircle,draw] (q0i2) at (1.8,2) {};
    \node[parcircle,draw] (q0i3) at (0,0.9) {};

    \node[parsquare,draw] (q1) at (-1.8,0.9) {$a_0,\underline{b_1},c_0$};
    \node[parcircle,draw] (q1i0) at (-3.4,0.9) {};
    \node[parcircle,draw] (q1i1) at (-1.8,0) {};

    \node[parsquare,draw] (q2) at (1.8,0.9) {$a_0,b_0,\underline{c_1}$};
    \node[parcircle,draw] (q2i0) at (3.4,0.9) {};
    \node[parcircle,draw] (q2i1) at (1.8,0) {};

    \node[parsquare,draw] (q3) at (0,-0.3) {$(a_0,\underline{b_1},c_1)$};
    
     \node[parsquare,draw] (q4) at (2.9,-0.2) {$(a_0,b_1,\underline{c_1})$};

%    \path[->,transform canvas={xshift=-0.5mm}] (q0) edge node[left,yshift=2mm] {$(0,0)$} (q0i0);
    \path[->,transform canvas={xshift=-0.5mm}] (q0) edge  (q0i0);
    \path[->,paredge,transform canvas={xshift=0.5mm}] (q0i0) edge node[right,yshift=2mm] {$(1,0)$} (q0);
    \path[->,paredge,transform canvas={yshift=0.5mm}] (q0i1) edge node[above,xshift=-1mm] {$(1,0)$} (q0);
%    \path[->,transform canvas={yshift=-0.5mm}] (q0) edge node[below,xshift=-1mm] {$(0,0)$} (q0i1);
    \path[->,transform canvas={yshift=-0.5mm}] (q0) edge (q0i1);
    \path[->,dotted] (q0i1) edge node[left] {$(0,1)$} (q1);    
%    \path[->,transform canvas={yshift=-0.5mm}] (q0) edge node[below,xshift=1mm] {$(0,0)$} (q0i2);
    \path[->,transform canvas={yshift=-0.5mm}] (q0) edge (q0i2);
    \path[->,paredge,transform canvas={yshift=0.5mm}] (q0i2) edge node[above,xshift=1mm] {$(1,0)$} (q0);    
    \path[->,dotted] (q0i2) edge node[left] {$(0,1)$} (q2);
%    \path[->] (q0) edge node[left,xshift=0.5mm,yshift=0.5mm] {$(0,0)$} (q0i3);
    \path[->] (q0) edge (q0i3);
    \path[->,very thick, densely dotted] (q0i3) edge node[above] {$(0,1)$} (q1);
    \path[->,very thick, densely dotted] (q0i3) edge node[above] {$(0,1)$} (q2);
    \path[->,dotted] (q0i3) edge node[left,yshift=5mm] {$(0,1)$} (q3);

%    \path[->,transform canvas={yshift=0.5mm}] (q1) edge node[above] {$(0,0)$} (q1i0);
    \path[->,transform canvas={yshift=0.5mm}] (q1) edge (q1i0);
    \path[->,paredgee,bend left=55] (q1i0) edge node[above,sloped,xshift=-3mm] {$(1,0)$} (q0);
    \path[->,dotted,transform canvas={yshift=-0.5mm}] (q1i0) edge node[below] {$(0,1)$} (q1);
%    \path[->,transform canvas={xshift=-0.5mm}] (q1) edge node[left] {$(0,0)$} (q1i1);
    \path[->,transform canvas={xshift=-0.5mm}] (q1) edge (q1i1);
    \path[->,paredgee] (q1i1) edge node[below,xshift=-4mm,sloped] {$(1,0)$} (q2);
    \path[->,dotted,transform canvas={xshift=0.5mm}] (q1i1) edge node[right] {$(0,1)$} (q1);
    \draw[->,dotted,-Triangle] (q1i1) |- node[below] {$(0,1)$} ([xshift=-2mm]q3.west);

%    \path[->,transform canvas={yshift=0.5mm}] (q2) edge node[above] {$(0,0)$} (q2i0);
    \path[->,transform canvas={yshift=0.5mm}] (q2) edge (q2i0);
    \path[->,paredgee,bend right=55] (q2i0) edge node[above,sloped,xshift=3mm] {$(1,0)$} (q0);
    \path[->,dotted,transform canvas={yshift=-0.5mm}] (q2i0) edge node[below] {$(0,1)$} (q2);
%    \path[->,transform canvas={xshift=-0.5mm}] (q2) edge node[left] {$(0,0)$} (q2i1);
    \path[->,transform canvas={xshift=-0.5mm}] (q2) edge (q2i1);
    \path[->,paredgee] (q2i1) edge node[below,xshift=4mm,sloped] {$(1,0)$} (q1);
    \path[->,dotted,transform canvas={xshift=0.5mm}] (q2i1) edge node[right] {$(1,0)$} (q2);
%    \draw[->,dotted,-Triangle] (q2i1) |- ([xshift=2mm]q3.east);
	\draw[->,dotted,-Triangle] (q2i1) |- node[below] {$(0,1)$} (q4);

    \node (dot) at (-1.8,3.1) {$\bullet$};
%    \path[dotted,->] (bot) edge[loop right] node[right=1pt] {\tiny $\textsf{true},1$} (dot);
    \draw[dotted,->] (q0i0) edge (dot);
    \draw[dotted,->] (q0i1) edge (dot);
    \draw[dotted,->] (q0i2) edge (dot);
    \draw[paredge,->] (q0i3) edge (dot);
    \draw[dotted,->, bend left=45] (q1i0) edge (dot);
    \draw[dotted,->,bend left=60] (q1i1) edge (dot);
    \draw[dotted,->] (q2i0) edge (dot);
    \draw[dotted,->] (q2i1) edge (dot);    

\end{tikzpicture}
}
\scalebox{0.75}{
\begin{tikzpicture}[scale=1.5,font={\tiny}]
    \node[parsquare,draw] (q0) at (0,2) {$a_0,\underline{b_0},c_0$};
    \node[parcircle,draw] (q0i0) at (0,3.1) {};
    \node[parcircle,draw] (q0i1) at (-1.8,2) {};
    \node[parcircle,draw] (q0i2) at (1.8,2) {};
    \node[parcircle,draw] (q0i3) at (0,0.9) {};

    \node[parsquare,draw] (q1) at (-1.8,0.9) {$a_0,\underline{b_1},c_0$};
    \node[parcircle,draw] (q1i0) at (-3.4,0.9) {};
    \node[parcircle,draw] (q1i1) at (-1.8,0) {};

    \node[parsquare,draw] (q2) at (1.8,0.9) {$a_0,b_0,\underline{c_1}$};
    \node[parcircle,draw] (q2i0) at (3.4,0.9) {};
    \node[parcircle,draw] (q2i1) at (1.8,0) {};

    \node[parsquare,draw] (q3) at (0,-0.3) {$(a_0,\underline{b_1},c_1)$};
    
     \node[parsquare,draw] (q4) at (2.9,-0.2) {$(a_0,b_1,\underline{c_1})$};

%    \path[->,transform canvas={xshift=-0.5mm}] (q0) edge node[left,yshift=2mm] {$(0,0)$} (q0i0);
    \path[->,transform canvas={xshift=-0.5mm}] (q0) edge  (q0i0);
    \path[->,paredge,transform canvas={xshift=0.5mm}] (q0i0) edge node[right,yshift=2mm] {$(1,0)$} (q0);
    \path[->,paredge,transform canvas={yshift=0.5mm}] (q0i1) edge node[above,xshift=-1mm] {$(1,0)$} (q0);
%    \path[->,transform canvas={yshift=-0.5mm}] (q0) edge node[below,xshift=-1mm] {$(0,0)$} (q0i1);
    \path[->,transform canvas={yshift=-0.5mm}] (q0) edge (q0i1);
    \path[->,dotted] (q0i1) edge node[left] {$(0,1)$} (q1);    
%    \path[->,transform canvas={yshift=-0.5mm}] (q0) edge node[below,xshift=1mm] {$(0,0)$} (q0i2);
    \path[->,transform canvas={yshift=-0.5mm}] (q0) edge (q0i2);
    \path[->,paredge,transform canvas={yshift=0.5mm}] (q0i2) edge node[above,xshift=1mm] {$(1,0)$} (q0);    
    \path[->,dotted] (q0i2) edge node[left] {$(0,1)$} (q2);
%    \path[->] (q0) edge node[left,xshift=0.5mm,yshift=0.5mm] {$(0,0)$} (q0i3);
    \path[->] (q0) edge (q0i3);
    \path[->,paredgee] (q0i3) edge node[above] {$(0,1)$} (q1);
    \path[->,paredgee] (q0i3) edge node[above] {$(0,1)$} (q2);
    \path[->,dotted] (q0i3) edge node[left,yshift=5mm] {$(0,1)$} (q3);

%    \path[->,transform canvas={yshift=0.5mm}] (q1) edge node[above] {$(0,0)$} (q1i0);
    \path[->,transform canvas={yshift=0.5mm}] (q1) edge (q1i0);
    \path[->,paredgee,bend left=55] (q1i0) edge node[above,sloped,xshift=-3mm] {$(1,0)$} (q0);
    \path[->,dotted,transform canvas={yshift=-0.5mm}] (q1i0) edge node[below] {$(0,1)$} (q1);
%    \path[->,transform canvas={xshift=-0.5mm}] (q1) edge node[left] {$(0,0)$} (q1i1);
    \path[->,transform canvas={xshift=-0.5mm}] (q1) edge (q1i1);
    \path[->,paredgee] (q1i1) edge node[below,xshift=-4mm,sloped] {$(1,0)$} (q2);
    \path[->,dotted,transform canvas={xshift=0.5mm}] (q1i1) edge node[right] {$(0,1)$} (q1);
    \draw[->,dotted,-Triangle] (q1i1) |- node[below] {$(0,1)$} ([xshift=-2mm]q3.west);

%    \path[->,transform canvas={yshift=0.5mm}] (q2) edge node[above] {$(0,0)$} (q2i0);
    \path[->,transform canvas={yshift=0.5mm}] (q2) edge (q2i0);
    \path[->,paredgee,bend right=55] (q2i0) edge node[above,sloped,xshift=3mm] {$(1,0)$} (q0);
    \path[->,dotted,transform canvas={yshift=-0.5mm}] (q2i0) edge node[below] {$(0,1)$} (q2);
%    \path[->,transform canvas={xshift=-0.5mm}] (q2) edge node[left] {$(0,0)$} (q2i1);
    \path[->,transform canvas={xshift=-0.5mm}] (q2) edge (q2i1);
    \path[->,paredgee] (q2i1) edge node[below,xshift=4mm,sloped] {$(1,0)$} (q1);
    \path[->,dotted,transform canvas={xshift=0.5mm}] (q2i1) edge node[right] {$(1,0)$} (q2);
%    \draw[->,dotted,-Triangle] (q2i1) |- ([xshift=2mm]q3.east);
	\draw[->,dotted,-Triangle] (q2i1) |- node[below] {$(0,1)$} (q4);

    \node (dot) at (-1.8,3.1) {$\bullet$};
%    \path[dotted,->] (bot) edge[loop right] node[right=1pt] {\tiny $\textsf{true},1$} (dot);
    \draw[dotted,->] (q0i0) edge (dot);
    \draw[dotted,->] (q0i1) edge (dot);
    \draw[dotted,->] (q0i2) edge (dot);
    \draw[dotted,->] (q0i3) edge (dot);
    \draw[dotted,->, bend left=45] (q1i0) edge (dot);
    \draw[dotted,->,bend left=60] (q1i1) edge (dot);
    \draw[dotted,->] (q2i0) edge (dot);
    \draw[dotted,->] (q2i1) edge (dot);    

\end{tikzpicture}
}
\end{center}
\caption{\footnotesize 
The arena after one exploration step as show in Fig.~\protect\ref{fig:pg-ex-3} but extended with the node $\bullet$.
%As in Fig.~\protect\ref{fig:pg-solve-1}, we have to decide if $\plr=\protect\pctrl$ wins $(a_0,\underline{b_0},c_0)$.
To simplify presentation, we have removed the node $\bot$ as $\protect\pctrl$ will always rather play to $\bullet$; we also have dropped the neutral edge weights i.e.\ every unlabelled edge is implicitly labeled by $(0,0)$.
Top figure: Here, we are given the strategy we computed in Fig.~\protect\ref{fig:pg-solve-1} as initial strategy for $\protect\pctrl$; 
for every node added by the exploration step, we extend the strategy by letting $\protect\pctrl$ play to $\bullet$ again.
We identify as improvements the thick dotted edges.
This gives us the strategy shown in the figure in the middle where the edges in green have been inherited from the initial strategy, while the edges in blue are the improvement identified in the preceding step of the strategy iteration.
W.r.t.\ this strategy we only identify one further improvement which leads to the final strategy shown in the lower figure that wins the node $(a_0,\underline{b_0},c_0)$.
}
\label{fig:pg-solve-2}
\end{figure}
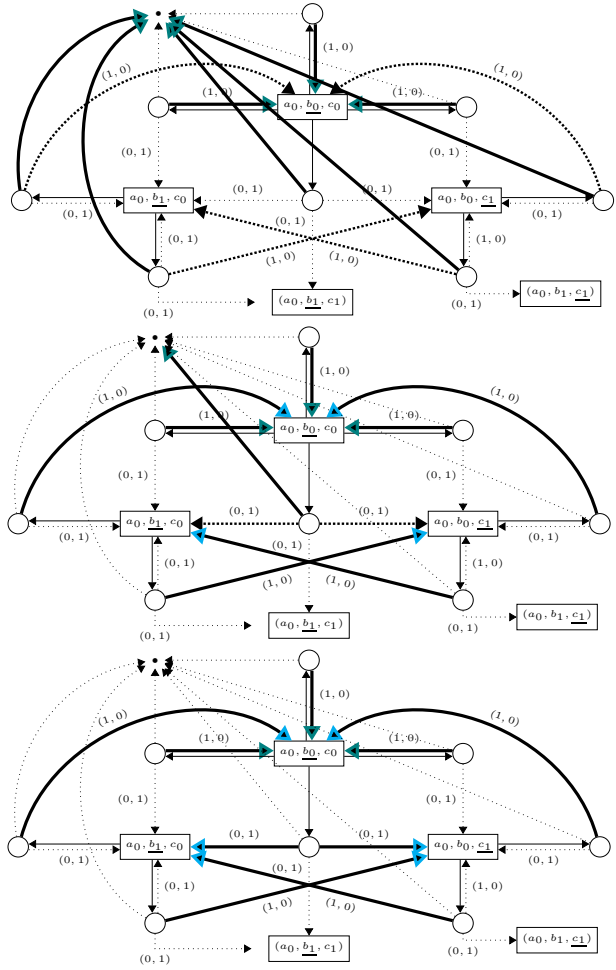

Computation of the sup-inf-distances in $\cA^\bullet$ is complicated by the existence of both positive and negative cycles (w.r.t.\ the interpretation of $\N_0^C$ as numerals).
For this reason, we use strategy iteration, i.e.\ we construct a sequence of strategies for $\plr$ where each strategy only allows for positive cycles, while negative cycles are prevented by playing to $\bullet$:
Assume $\plr$ uses a (nondeterministic memoryless) strategy $\kappa$ (i.e.\ $\emptyset\neq\kappa(v) \subseteq vE$ for all $v\in V_P$) s.t.\ in the accordingly restricted arena $\cA^\bullet_\kappa$ (i.e.\ $\plr$ may only move to a successor in $\kappa(v)$ at every node $v\in V_{\plr}$) s.t. no negative cycles exist anymore (e.g.\ consider $\kappa = V_P\times\{\bullet\}$); then the sup-inf-distance for every node can be easily computed using fixed-point iteration just as in the case of the standard attractor computation using some variant of the Bellmann-Ford algorithm (in order to identify infinite positive sup-inf-distance).
 Let $d_{\kappa}(v)$ denote the sup-inf-distance of node $v$ in $\cA^\bullet_{\kappa}$. For every node $v$ in $B\cup \{\bullet\}$, we define $d_{\kappa}(v)=0$; for the nodes $\bot$ and $\top$ we predefine $d_{\kappa}$ accordingly; for every $v$ controlled by $\opp$ we have the unrestricted optimality equation $d_{\kappa}(v) = \min^{\pord}\{ \gamma(v,w)+d_{\kappa}(w) \mid w\in vE\}$, while for $v$ controlled by $\plr$ we have the $\kappa$-restricted optimality equation $d_{\kappa}(v)=\max^{\pord}\{ \gamma(v,w)+ d_{\kappa}(w) \mid w\in\kappa(v)\}$ (with $\gamma(v,w)$ the weight induced by the colour of the edge from $v$ to $w$). We call any nondeterministic strategy $\kappa'$ an improvement of $\kappa$ if (1) $\kappa'\neq \kappa$, (2) $\emptyset\neq \kappa'(v) \subseteq vE$ for all $v\in V_P$, (3) if $(v,w)\in \kappa'\setminus \kappa$, then $d_{\kappa}(v) < \gamma(v,w)+d_{\kappa}(w)$ (i.e.\ $v$'s sup-inf-distance can be improved by playing to $w$), and (4) if $(v,w)\in \kappa'\cap \kappa$, then $d_{\kappa}(v)=\gamma(v,w)+d_{\kappa}(w)$. \cite{DBLP:journals/corr/abs-0806-2923} shows that any such improvement $\kappa'$ does not introduce any negative cycles and that $d_{\kappa'} > d_{\kappa}$, i.e.\ the sup-inf-distances w.r.t.\ $\kappa'$ do not decrease for any node and strictly improves for at least one node.
 Thus, no strategy can be encountered twice. Finally, note that if $W_{\plr}\uplus W_{\opp}$ is the winning partition of the nodes $V$ of $\cA$ w.r.t.\ the min-parity-condition, then $\plr$ can use his (memoryless deterministic) winning strategy from the parity game to ensure infinite sup-inf-distance on $W_{\plr}$ also in $\cA^\bullet$, while $\opp$ can use his winning strategy from the parity game to ensure at least finite sup-inf-distance in $\cA^\bullet$.
Hence, eventually the strategy iteration will terminate in a strategy that guarantees infinite sup-inf-distance on exactly $W_P$.

In our implementation, we make use of the fact that the fixed-point iteration used for solving the optimality equations stated above can be easily parallelised in order to make use of modern multi-core CPUs (or even GPUs~\cite{DBLP:conf/atva/MeyerL16}).

\subsubsection{Exploring the Boundary}\label{sec:boundary}

$\oexpl$ selects nodes from the boundary $B$ that should be further explored, i.e., where successors should be computed. 
After each expansion $B$ is recomputed such that it contains the nodes which successors have not yet been explored.
We instantiate $\oexpl$ with two different approaches.
One is based on \emph{breadth-first search} exploration, and one on a \emph{priority queue} ordering states in the boundary by some \emph{quality score}.
For each method, we then have an additional variant that \emph{filters} the states that are currently needed to determine the winner of the arena.
In total, this results in the following four methods.

\begin{itemize}
    \item $\oexplbfs$ (breadth-first search (BFS) exploration):
        In the initial implementation the algorithm explored and constructed the parity game using a breadth-first-search (BFS). This approach helps to ignore parts of the game that are far away from the initial state and not decisive for winning the game, however it also explores states that are close to the initial state, but are irrelevant with the currently computed strategy.

Conceptually, we use the exploration function that picks states from $B$ with a minimal distance from the initial state $q_0$ in each step. Define $\oexplbfs$ by:
\begin{align*}
\oexplbfs(\alpha, \vctrl, \venv, E, \chi, B, q_0, \parity, \sigma, \tau) := \argmin_{b \in B} \dist(\vctrl \cup \venv, E, q_0, b) \\
\dist(V, E, q_0, q) := \min\{ \abs{\pi} \mid \pi \text{ is a path from $q_0$ to $q$ in $(V, E)$} \}
\end{align*}
In the implementation, instead of recomputing the minimal distances in each step, we select the next state from the boundary from a worklist queue.

    \item $\oexplbfsclear$ (BFS exploration with strategy-based worklist filtering):
This exploration strategy is a variant of $\oexplbfs$.
The worklist is still populated in a BFS way, but we only keep states that we know are needed to determine if the initial state is winning or losing.
These are states which are reachable through some path not blocked by an already winning or losing state.
We define the exploration function $\oexplbfsclear$ as follows:
\begin{align*}
    \oexplbfsclear&(\alpha, \vctrl, \venv, E, \chi, B, q_0, \parity, \sigma, \tau) := \\
    &\oexplbfs(\alpha, \vctrl, \venv, E, \chi, \filter(B), q_0, \parity, \sigma, \tau) \\
    \filter(B) &:= \{ b \in B \mid \exists\;\text{path $\pi$ from $q_0$ to $b$ s.t. }\forall q \in \pi : \winner(q) = \symunknown\, \} \\
    \winner(q) &:=
    \begin{cases}
        \pctrl & \text{if $\owin(\vctrl, \venv, E, \chi, B, q, \pctrl, \parity, \sigma)$ returns true} \\
        \penv & \text{if $\owin(\vctrl, \venv, E, \chi, B, q, \penv, \overline{\parity}, \tau)$ returns true} \\
        \symunknown & \text{otherwise}
    \end{cases}
\end{align*}
We reuse the results of $\owin$ already computed in the main algorithm.
However, in contrast to $\oexplbfs$, we need to check for existence of a path from $q_0$
to all states $b \in B$ without already won states, which we do using a single linear-time search in each iteration.

    \item $\oexplpq$ (priority queue exploration based on scores):
This exploration method is based on \emph{quality scores} (to be defined in the next section) assigned to each state
in the boundary.
The idea is that a high score means this state is a promising state for $\pctrl$ to win the game from the initial state,
while a low score is a promising state for $\penv$ to win the game.
As we initially do not know if the specification is realisable or unrealisable, and thus do not know
for which player the initial state is winning, this method explores both states with high and low scores
simultaneously. We assign scores not to states, but to edges, to incorporate information from colours
and updates in the LAR. Let $\oscore(\alpha, q, a, q')$ be a function assigning a score to a given edge $(q,a,q')$
of the parity automaton $\cP(\alpha)$.
With an intermediate function $s(b)$ assigning to a state $b \in B$ all scores of incoming edges,
we define the exploration method $\oexplpq$ by:
\begin{align*}
    \oexplpq&(\alpha, \vctrl, \venv, E, \chi, B, q_0, \parity, \sigma, \tau) \coloneqq \\
    &\left( \argmin_{b \in B} (\min s(b)) \right) \cup \left( \argmax_{b \in B} (\max s(b)) \right) \\
    s(b) &\coloneqq \{ \oscore(q,i \cup o,b) \mid (q,I,q'),(q',O,b) \in E,\, i \in I,\, o \in O \}
\end{align*}
We implement this method with a double-ended priority queue, in which states are
inserted with their respective minimal and maximal score of an incoming edge upon discovery.

    \item $\oexplpqclear$ (priority queue exploration with strategy-based worklist filtering):
        Finally, we can combine the exploration using scores with the filtering function $\filter$ from $\oexplbfsclear$,
        yielding the method $\oexplpqclear$ defined by:
        \begin{align*}
            \oexplpqclear&(\alpha, \vctrl, \venv, E, \chi, B, q_0, \parity, \sigma, \tau) := \\
            &\oexplpq(\alpha, \vctrl, \venv, E, \chi, \filter(B), q_0, \parity, \sigma, \tau)
        \end{align*}
\end{itemize}

\subsubsection{Quality Scores}

We now describe how to compute scores by $\oscore(\decompose_\varphi, q,a,q')$ for an edge $(q,a,q')$ of the DPA $\cP(\decompose_\varphi)$
leading to a boundary state $q'$.
We first define a function $\score$, which takes the
acceptance-type annotated formula $\decompose_\varphi$ and an edge $(q,a,q')$ of $\cP(\decompose_\varphi)$,
and returns a \emph{weighted score} $(w,s) = \score(\decompose_\varphi,q,a,q')$.
We always assign the score value $s$ such that $s = 0$ if $q' = \bot$, $s = 1$ if $q' = \top$ and $0 < s < 1$ otherwise.
The function $\score(\decompose_\varphi,q,a,q')$ computes $(w,s)$ recursively on $\decompose_\varphi$, similarly to the LAR-product construction,
and we define it by case distinction on $\decompose_\varphi$ and $q$.

\begin{itemize}
    \item If $q = \bot$, return $w = 1$ and $s = 0$.
    \item If $q = \top$, return $w = 1$ and $s = 1$.
    \item In the base case for $\mathcal{W}(\varphi), \mathcal{B}(\varphi), \mathcal{C}(\varphi)$ and $\mathcal{P}(\varphi)$, return $w = 1$
        and compute $s$ depending on the type of automaton. 
        \begin{itemize}
            \item If we encounter $\mathcal{W}(\varphi)$, then $\varphi \in \ltlfum \cup \ltlgrw$. By using the construction of \cite{DBLP:conf/lics/EsparzaKS18} the state $q'$ is in fact a Boolean formula over modal operators treated as variables. Let $\mathcal{V}$ be the set of variables and let $\mathcal{M}$ be the set of satisfying assignments of $q'$. We then set $s = |\mathcal{M}| / 2^{|\mathcal{V}|}$. Using this definition, we assign to the state $\true$ (which corresponds to $\top$) $s = 1$ and to the state $\false$ (which corresponds to $\bot$) $s = 0$.
            \item If we encounter $\mathcal{B}(\varphi)$, then we apply the construction of \cite{DBLP:conf/cav/SickertEJK16} and the state $q' = (p, p')$ is a tuple of two Boolean formulas $p$ and $p'$. We then construct $p \wedge p'$ and compute the scoring for $\mathcal{W}(p \wedge p')$.
            \item If we encounter $\mathcal{C}(\varphi)$, then we apply the same approach as in $B(\varphi)$, since these automata are obtained by complementing Büchi automata and have an identical structure besides the acceptance condition.
            \item If we encounter $\mathcal{P}(\varphi)$, we bail and return $s = \frac{1}{2}$.
        \end{itemize}
    \item In the case $\decompose_\varphi = \mathcal{X}_\land(\mathcal{X}_1,\ldots,\mathcal{X}_n)$ and $\decompose_\varphi = \mathcal{X}_\lor(\mathcal{X}_1,\ldots,\mathcal{X}_n)$, we first compute the child scores $(w_i,s_i) = \score(\mathcal{X}_i,q,a,q')$.
        Then, for each $\mathcal{X}_i$, we update $w_i$ and $s_i$ as follows:
    \begin{itemize}
        \item If $\decompose_\varphi = \mathcal{X}_\land$, set $w_i \leftarrow w_i \cdot \log_{\nicefrac{1}{2}} s_i$. This gives score values close to $0$ an increased weight,
            which could make the whole conjunction false.
        \item If $\decompose_\varphi = \mathcal{X}_\lor$, set $w_i \leftarrow w_i \cdot \log_{\nicefrac{1}{2}} (1 - s_i)$. This gives score values close to $1$ an increased weight,
            which could make the whole disjunction true.
    \item If $\decompose_\varphi = \mathcal{B}_\land(\mathcal{B}_1, \ldots, \mathcal{B}_n)$, or $\decompose_\varphi = \mathcal{C}_\lor(\mathcal{C}_1, \ldots, \mathcal{C}_n)$, then $\mathcal{X}_i = \mathcal{B}$ or $\mathcal{X}_i = \mathcal{C}$ and $q = (q'',r)$ and $q' = (q''',r')$ for some round-robin counters $r,r'$. Now if the child caused the round-robin counter to increase, i.e. we have $r \le i < r'$, then we set $w_i \leftarrow 2 \cdot w_i$ and update the score.
            If $\mathcal{X}_i = \mathcal{B}$, set $s_i \leftarrow \frac{3+s_i}{4}$ to increase the score, and if
            $\mathcal{X}_i = \mathcal{C}$, set $s_i \leftarrow \frac{s_i}{4}$ to decrease the score.
      \item If either $\decompose_\varphi = \mathcal{P}_\land(\mathcal{X}_1, \mathcal{X}_2)$, $\mathcal{X}_i = \mathcal{P}$, and $\mathcal{X}_{3-i} = \mathcal{B}$,
            or $\decompose_\varphi = \mathcal{P}_\lor(\mathcal{X}_1, \mathcal{X}_2)$, $\mathcal{X}_i = \mathcal{P}$, and $\mathcal{X}_{3-i} = \mathcal{C}$, then $q = (q'',c)$ and $q' = (q''',c')$ for some minimal colour memory
            values $c,c'$. Let $\parity$ be the parity of the DPA $\cP(\decompose_\varphi)$.
            Now if $\cP(\mathcal{X}_i)$ caused the memory to decrease to a value with parity $\parity$, we increase the score, and if it decreased it to a value with different parity, we decrease the score.
            Therefore if $c' < c$ and $c' \equiv_2 \parity$, then set
            $w_i \leftarrow 2 \cdot w_i$ and $s_i \leftarrow \frac{3+s_i}{4}$,
            and if $c' < c$ and $c' \equiv_ 2 \overline{\parity}$, then set
            $w_i \leftarrow 2 \cdot w_i$ and $s_i \leftarrow \frac{s_i}{4}$.
    \end{itemize}
    Finally, we return $w \coloneqq \sum_{i=1}^n w_i$ and $s \coloneqq \left(\sum_{i=1}^n w_i \cdot s_i\right) \mathbin{/} w$.
    \item In the case $\mathcal{X}_\leftrightarrow(\mathcal{X}_1,\mathcal{X}_2)$, we also first compute the child scores
    $(w_i,s_i) = \score(\mathcal{X}_i,q,a,q')$ for $i \in \{1,2\}$, and then update each $w_i$ and $s_i$ similarly to conjunction and disjunction:
        \begin{itemize}
            \item If $0 < s_i < 1$, set $w_i \leftarrow w_i \cdot \max(\log_{\nicefrac{1}{2}} s_i, \log_{\nicefrac{1}{2}}(1 - s_i))$.
            \item If $\decompose_\varphi = \mathcal{P}_\leftrightarrow(\mathcal{X}_1, \mathcal{X}_2)$ and $\cP(\mathcal{X}_i)$ is the child for which we store the colours in memory,
            then $q = (q'',c)$ and $q' = (q''',c')$ for some minimal colour memory values $c,c'$ of $\cP(\mathcal{X}_i)$.
            Let $\parity$ be the parity of the DPA $\cP(\decompose_\varphi)$.
            As for $\mathcal{P}_\land$ and $\mathcal{P}_\lor$, if $c' < c$ and $c' \equiv_2 \parity$, then set
            $w_i \leftarrow 2 \cdot w_i$ and $s_i \leftarrow \frac{3+s_i}{4}$,
            and if $c' < c$ and $c' \equiv_2 \overline{\parity}$, then set
            $w_i \leftarrow 2 \cdot w_i$ and $s_i \leftarrow \frac{s_i}{4}$.
    \end{itemize}
    Then we return $w \coloneqq \sum_{i=1}^n w_i$ and $s \coloneqq \left(\sum_{i=1}^n w_i \cdot s_i\right) \mathbin{/} w$.
\end{itemize}

We then define $\oscore(\decompose_\varphi,q,a,q') \coloneqq s$ where $(w,s) = \score(\decompose_\varphi,q,a,q')$.
The value $w$ is only necessary to normalise intermediate scores due to the recursive definition of $\score$.

\subsection{Controller Extraction}
\label{sec:extraction}

\subsubsection{Mealy machine}

When we determine that $\pctrl$ has a winning strategy $\sigma$ from $q_0$, we can extract
a controller from $\sigma$ that ensures realisation of the specification.

We use an \emph{incompletely specified} Mealy machine,
where some outputs might not be specified and could be instantiated either way.
This allows further minimisation and more compact representations by a circuit.
Given an input/output partition $\ap = \apin \uplus \apout$, a Mealy machine is a tuple $M = (Q, q_0, \delta, \lambda)$
where $Q$ is a finite set of states,
$q_0 \in Q$ is the initial state,
$\delta : Q \times 2^\apin \to Q$ is the transition function and
$\lambda: Q \times 2^\apout \to \{0,1,\symunknown\}^\apout$ is the output function,
where $\symunknown$ stands for an unspecified output. The output
can be given by a Boolean product term, where missing variables are unspecified.

Let $(\vctrl, \venv, E)$ be the parity game arena
where $\pctrl$ wins from $q_0$ with the strategy $\sigma$.
We use  $Q \coloneqq \{ q \in \venv \mid \owin(\vctrl, \venv, E, \chi, B, q, \pctrl, \parity, \sigma) \}$
as the set of states.
For the transition function, we define $\delta(q, i) \coloneqq q'$ by
choosing some $q'$ where $((q,I),O,q')  \in \sigma$ for some $I \subseteq \alin$ with $i \in I$ and any $O \subseteq \alout$.
By construction, and as $\sigma$ is a winning strategy for all $q \in Q$, such a $q'$ always exists.
However, there may be multiple applicable $q'$.
We construct $Q$ and $\delta$ iteratively, starting from $Q \leftarrow \{q_0\}$, and try to choose
for every $q \in Q$ and $i \in 2^\apin$ a $q' = \delta(q,i)$ such that $q' \in Q$, if possible, and otherwise extend $Q \leftarrow Q \cup \{q'\}$.
As a secondary heuristic, we try to choose a successor that gives the most flexibility in choosing the output,
i.e. a $q'$ such that $\sum \{ |O| \mid \exists I: i \in I \land ((q,I),O,q') \in \sigma \}$ is maximal.

For the output function, let $\delta(q,i) = q'$ be an edge of the Mealy machine;
then take a minimal prime implicant $o'$ of the Boolean formula over $\apout$ that encodes the set
$\{ o \mid \exists I,O : i \in I \land o \in O \land (q,I),O,q') \in \sigma \}$, and define $\lambda(q, i) \coloneqq o'$.
This again exploits non-determinism of the strategy $\sigma$.

\begin{example}
Using the winning strategy for the simple arbiter specification in Fig.~\ref{fig:pg-ex-3},
we obtain $Q = \{ (a_0,\underline{b_0},c_0), (a_0,\underline{b_1},c_0), (a_0,b_0,\underline{c_1}) \}$
as states of the Mealy machine.
For $\delta((a_0,\underline{b_0},c_0), r_1 r_2)$, we may choose any of the other two states
as a successor. If we choose $(a_0,\underline{b_1},c_0)$, we get the output
$o = \lambda((a_0,\underline{b_0},c_0), r_1 r_2) = \overline{g_1} g_2$, specifying
$o(g_1) = 0$ and $o(g_2) = 1$.
For the output $\lambda((a_0,\underline{b_0},c_0), \overline{r_1}\, \overline{r_2})$,
we may choose one of the two min-terms $\overline{g_1}$ or $\overline{g_2}$ as an output.
If we choose $o = \overline{g_1}$, then $o(g_1) = 0$, but the output $o(g_2) = \symunknown$ is unspecified.
An implementation is then free to choose $o(g_2) = 0$ or $o(g_2) = 1$.
\end{example}

This incompletely specified Mealy machine can optionally be \emph{minimised}.
We use the exact minimisation algorithm from~\cite{DBLP:conf/iccad/0002R15},
which in turn uses a SAT solver.
While this problem is harder than minimisation of fully specified Mealy machines,
it can also result in smaller machines.

\subsubsection{Controller as BDD or AIG}\label{sec:encoding}

While we can directly output the controller as a Mealy machine,
we can also encode it as a \emph{binary decision diagram (BDD)} or
\emph{and-inverter graph (AIG)}, representing a circuit.
For this, we need to encode the transition function $\delta$ and output function $\lambda$
of the Mealy machine as a BDD or AIG.
Both of these representations use decisions over binary variables.
The inputs $i \in 2^\apin$ and outputs $o \in \{0,1,\symunknown\}^\apout$ are already binary vectors
after one resolves the unspecified $\symunknown$ outputs.
However, one needs to choose a binary encoding of the state space $Q$.
Here we offer two options:

First, we can use an \emph{unstructured encoding} $\lunstruct$.
We simply enumerate the states $Q = \{q_0,\ldots,q_n\}$
and use the encoding function
$\lunstruct: Q \to 2^{\lceil \log_2 |Q| \rceil}$
with $\lunstruct(q_i) = i$ which maps each state to the binary encoding of its number.

Second, we can use the shape of the states for a \emph{structured encoding} $\lstruct$.
As a state $q$ is a vector $(q_1,\ldots,q_n)$, we can encode each
component separately into a binary vector.
Assume that for each component $1 \le i \le n$, the states $q_i$
are numbered from $0$ to $|Q_i|-1$.
Then we use the encoding function
$\lstruct: Q \to 2^{\sum_{i=1}^n \lceil \log_2 |Q_i| \rceil}$,
with
\[
    l(q_1, \ldots, q_n) \coloneqq
    \sum_{i=1}^n q_i \cdot 2^{\sum_{j=1}^{i-1}\lceil \log_2 |Q_i| \rceil}
\]
which concatenates the binary encoding of the state number in each component.
Note that this also includes additional memory information such as the
round-robin counter or minimal colour memory.

\begin{example}
For the Mealy machine obtained from the strategy in Fig.~\ref{fig:pg-ex-3},
we have as states $Q = \{ (a_0,\underline{b_0},c_0), (a_0,\underline{b_1},c_0), (a_0,b_0,\underline{c_1}) \}$.
With the unstructured encoding for $Q=\{q_0,q_1,q_2\}$, we get the binary state encodings
$\lunstruct(Q) = \{ 00_2, 01_2, 10_2 \}$.
If we represent the product states as numbers with $q_i = i$ for $q \in \{a,b,c\}$
and the round-robin counter by $r \in \{0,1\}$,
then after applying the structured encoding function, we get the state encodings
$\lstruct(Q) = \{ 0000_2, 0100_2, 0011_2 \}$.
\end{example}

We also minimise the controller for these output formats.
When constructing a BDD, we minimise it using the CUDD library~\cite{CUDD}
by reordering the variables.
When constructing an AIG, we first construct a BDD and minimise it,
and then construct the AIG from the BDD.
Afterwards, we minimise the AIG using functionality from the
ABC library~\cite{DBLP:conf/cav/BraytonM10}.

When using the encoding function, it can sometimes also be more effective
to \emph{not} minimise the Mealy machine before, as this can destroy the
structure from the product state.
Structured encoding can also sometimes increase the size of a circuit.
We offer an option to construct a circuit using the three
following combinations in parallel and then return the smallest circuit:
Mealy minimisation and unstructured encoding,
no minimisation and structured encoding, and no minimisation and unstructured encoding.

\section{Experimental Evaluation}\label{cha:experimental_evaluation}

The three main research questions we want to answer in this section are:

\begin{itemize}
	\item \emph{RQ1}: How does \toolname{} compare to existing tools? Specifically, we analyse:
	\begin{itemize}
		\item Number of instances correctly identified to be realisable.
		\item Number of instances for which a correct controller was synthesised.
		\item Circuit size of the constructed controller.
		\item Performance with increasing alphabet size.
	\end{itemize}
	\item \emph{RQ2}: What is the difference in performance of the proposed exploration strategies (\textsf{bfs} and \textsf{pq}, with and without \textsf{filtering})? Specifically, we analyse:
    \begin{itemize}
		\item Number of constructed states.
		\item Runtime.
	\end{itemize}
	\item \emph{RQ3}:  What is the difference in size of the proposed circuit encoding strategies (\textsf{structured} and \textsf{unstructured}, with and without \textsf{minimisation})? Specifically, we analyse:
    \begin{itemize}
   		\item Circuit size.
   	\end{itemize}
\end{itemize}

\paragraph{Experimental Design.}

We approach all three research questions by evaluating the different tools and configurations on the specifications from the TLFS/LTL-track of the \SC{} competition\footnote{\url{http://www.syntcomp.org/}}, which subsumes all benchmarks of \SCet{}~\cite{DBLP:journals/corr/abs-1904-07736}. To the best of our knowledge this dataset is the most complete set of LTL specifications for synthesis stemming from a wide range of different applications.
These include industrial examples such as the AMBA AHB arbiter~\cite{Jobstmann07,DBLP:journals/sttt/GodhalCH13,DBLP:journals/corr/BloemJK14a},
and case studies for hardware controller synthesis~\cite{DBLP:conf/haskell/Finkbeiner0PS19,XXXX:conf/fmcad/G19}.
For more details, see previous \syntcomp{} competition reports~\cite{DBLP:journals/corr/JacobsBBK0KKLNP16,DBLP:journals/corr/abs-1711-11439,DBLP:journals/corr/abs-1904-07736}.
The \SC{} set of specifications contains in total 434 LTL synthesis specifications, of which 337 are realisable and 97 are unrealisable. All experiments were run on a server with an Intel E5-2630 v4 clocked at 2.2\,GHz (boost disabled, 40 cores). We imposed a memory limit of 100\,GB (as in \SC{}) and a wall-clock time limit of one hour for each specification.

In Section~\ref{subsec:comp-ltlsynt} we address \emph{RQ1} and evaluate overall performance on the benchmark set and compare against \ltlsynt{}, \bosy{} and a previous version of \toolname{}. In Section~\ref{subsec:comp-expl} (\emph{RQ2}), we compare the different exploration strategies, and in Section~\ref{subsec:comp-min} (\emph{RQ3}) different construction approaches for the circuit.

\paragraph{Independent Evaluations.}

Since the first release of \toolname{} \cite{DBLP:conf/cav/MeyerSL18} independent researchers used, evaluated, and compared it to other tools.
At \SC{}, \toolname{} in its submitted version (19.07) again made first place in all categories in the LTL synthesis track. In two case studies \cite{DBLP:conf/haskell/Finkbeiner0PS19,XXXX:conf/fmcad/G19} by Finkbeiner et al.\ \toolname{} was used to synthesise controllers for small hardware devices; in the second case study \cite{XXXX:conf/fmcad/G19}, \toolname{} was also compared to \bosy{} and the not publicly available {\sc BoWSer} where \toolname{} was the only tool to solve all synthesis problems, and also performed best w.r.t.\ time and size of the obtained controller in almost all of the synthesis problems; interestingly, {\sc BoWSer} clearly outperformed \toolname{} in the specification \texttt{SensorSelector} where {\sc BoWSer} was roughly seven times faster than \toolname{} and also succeeds in finding a trivial controller with zero gates within 38 seconds while \toolname{} needed 280 seconds for returning a controller using 17 gates.

Finally, \cite{DBLP:conf/atva/EhlersA19} compares \toolname{} to {\sc GuiSynth}, a tool for synthesising code for graphical user interfaces. Here, {\sc GuiSynth} clearly outperforms \toolname{} in most benchmarks. As Ehlers and Adabala already noted this probably has to be contributed to the fact that {\sc GuiSynth} is specifically designed for the task at hand and the embedding into a standard LTL synthesis problem causes an exponential blow-up in the alphabet size.

\subsection{\emph{RQ1}: Comparison with previous version, with \ltlsynt{}, and with \bosy{}}\label{subsec:comp-ltlsynt}

\paragraph{Experimental Design.}

We run \toolname{} using the default exploration strategy $\oexplbfs$, AIG output and portfolio minimisation to find the smallest AIG.
We compare against \toolname{} in the version submitted to CAV 2018~\cite{DBLP:conf/cav/MeyerSL18} in April 2018.
This version also used the exploration strategy $\oexplbfs$, but did not have many of the improvements such as formula isomorphism detection, memoization,
symbolic representation of edges, or structured encoding.
Further, we compare our implementation with \bosy{} and \ltlsynt{} which achieved second\footnote{\toolname{} was ranked on the first place.} and third place in \SCet{} in the \enquote{synthesis quantity}-ranking.
We use \ltlsynt{} from the Spot library~\cite{DBLP:conf/atva/Duret-LutzLFMRX16,DBLP:journals/corr/abs-1711-11439} in the version 2.8.1 from July 2019, with parameter \texttt{-{}-algo=ds},
and against \bosy{}~\cite{DBLP:conf/cav/FaymonvilleFT17} in the newest version available as of July 2019, with parameter \texttt{-{}-optimize}.

\begin{table}
    \centering
    \begin{tabular}{lrrrrr}
        \toprule
         & & \multicolumn{1}{c}{\toolname{}} & \multicolumn{1}{c}{\toolname{}} & \multicolumn{1}{c}{\ltlsynt{}} & \multicolumn{1}{c}{\bosy{}} \\
         & max & \multicolumn{1}{c}{(19.07)} & \multicolumn{1}{c}{(CAV'18)} & \multicolumn{1}{c}{(2.8.1)} & \multicolumn{1}{c}{(July '19)} \\
        \midrule
        \textbf{Realisability} & 434  & \textbf{415}\,(29)   & 374\,(0) & 353\,(0) & 344\,(0) \\
        \textbf{Synthesis}     & 337  & \textbf{304}\,(28)   & 264\,(0) & 254\,(0) & 234\,(2) \\
        \textbf{Total Quality} & 674  & \textbf{571.66}      & 425.17   & 242.82   & 416.94   \\
        \textbf{Avg. Quality}  & 2.00 & \textbf{1.88}        &   1.61   &   0.96   &   1.78   \\
        \bottomrule
    \end{tabular}
    \caption{Overall results for \toolname{} compared with its previous version, \ltlsynt{} and \bosy{}.
        For \textbf{Realisability} and \textbf{Synthesis}, we give the number of solved instances (unique instances)
        and for \textbf{Quality} the total accumulated and the average points over solved synthesis instances.
    }\label{tab:experimental-results-comparison-overall}
\end{table}

We run each tool on each specification twice: once to check only realisability and in the realisable case once more to synthesise a controller in the AIGER format.
Table~\ref{tab:experimental-results-comparison-overall} gives the overall results.

The category \textbf{Realisability} counts the number of specifications for which realisability is correctly decided within the time limit, and
the category \textbf{Synthesis} counts the number of realisable specifications for which additionally a successfully verified controller is produced.
For this we verified the circuits with an additional time limit of one hour using the \tool{nuXmv} model checker~\cite{DBLP:conf/cav/CavadaCDGMMMRT14}
in version 1.1.1 with the \texttt{check\_ltlspec\_klive} routine.

The \textbf{Quality} rating compares the size of the solutions according to the \SC{} formula, where a tool gets
$\max(0, 2-\log_{10}\frac{n+1}{r+1})$ quality points
for each verified solution of size $n$ for a specification with reference size $r$.
The size of a solution is given by the number of and gates plus number of latches,
and as reference size we chose the smallest size of a verified solution produced by any of the four tools during our experiments.

Further, we list notable outliers in pairwise comparisons with other approaches. We compute for each pair of tools and each specification successfully solved by both tools the difference in order of magnitudes ($|\log_{10} \frac{x}{y}|$)\footnote{...or $|\log_{10} \frac{x + 1}{y + 1}|$ to compensate for circuits of size 0.} and select for each pair of tools the eight largest differences.

In order to study the impact of growing alphabet sizes we look at 4 parametrised benchmarks from \SC{} and measure the execution time for growing parameters. These parameters scale up the number of components and the size of the alphabet. Namely, we pick \texttt{arbiter (AMBA)}, \texttt{full\_arbiter\_enc}, \texttt{ltl2dba\_Q}, and \texttt{ltl2dba\_beta}.

\begin{table}
    \setlength{\tabcolsep}{4pt}
    \centering
    \scalebox{0.9}{
    \begin{tabular}{clrrrrrr}
        \toprule
        & & & & \multicolumn{1}{c}{\toolname{}} & \multicolumn{1}{c}{\toolname{}} & \multicolumn{1}{c}{\ltlsynt{}} & \multicolumn{1}{c}{\bosy{}} \\
        & Specification & $|\apin|$ & $|\apout|$ & \multicolumn{1}{c}{(19.07)} & \multicolumn{1}{c}{(CAV'18)} & \multicolumn{1}{c}{(2.8.1)} & \multicolumn{1}{c}{(July '19)} \\
        \midrule
        \multirow{8}{*}{\rotatebox[origin=c]{90}{\parbox{0.9cm}{\centering \toolname{}\\(19.07)} \emph{vs} \parbox{1.2cm}{\centering \toolname{}\\(CAV'18)}}}
& \texttt{simple\_arbiter\_10}                       &  10 &  10 &   \textbf{2.4}&        1603.2 &       818.0 &  \timeout \\
& \texttt{detector\_10}                              &  10 &   1 &   \textbf{0.8}&         486.4 &     \memout &     506.0 \\
& \texttt{prioritized\_arbiter\_8}                   &   9 &   9 &   \textbf{2.1}&        1048.6 &      2450.7 &  \timeout \\
& \texttt{ltl2dba\_C2\_10}                           &  10 &   1 &   \textbf{1.1}&         457.2 &     \memout &     474.0 \\
& \texttt{detector\_unreal\_10}                      &  10 &   1 &   \textbf{1.7}&         465.1 &     \memout &      13.1 \\
& \texttt{loadfull5}                                 &   6 &   5 &           5.0 &        1253.9 & \textbf{3.5}&      10.9 \\
& \texttt{amba\_decomposed\_lock\_8}                 &  17 &   1 &   \textbf{1.8}&         301.0 &        18.4 &       2.5 \\
& \texttt{ltl2dba\_theta\_8}                         &  10 &   1 &   \textbf{1.4}&         223.9 &     \memout &      96.5 \\
        \midrule
        \multirow{8}{*}{\rotatebox[origin=c]{90}{\parbox{0.9cm}{\centering \toolname{}\\(19.07)} \emph{vs} \parbox{1.2cm}{\centering \ltlsynt{}\\(2.8.1)}}}
& \texttt{ltl2dba\_C2\_8}                            &   8 &   1 &   \textbf{1.2}&          17.8 &      3443.6 &       7.1 \\
& \texttt{detector\_unreal\_8}                       &   8 &   1 &   \textbf{1.4}&          19.7 &      3320.2 &       1.5 \\
& \texttt{ltl2dpa22}                                 &   6 &   3 &   \textbf{1.2}&      \timeout &      2581.1 &      12.2 \\
& \texttt{prioritized\_arbiter\_8}                   &   9 &   9 &   \textbf{2.1}&        1048.6 &      2450.7 &  \timeout \\
& \texttt{prioritized\_...\_unreal1\_3\_8}                  &   4 &   4 &   \textbf{1.7}&      \timeout &      1686.7 &  \timeout \\
& \texttt{TorcsSteeringSmart}                        &   4 &   6 &          51.2 &          17.7 & \textbf{0.1}&       0.7 \\
& \texttt{torcs\_steering\_smart}                    &   4 &   6 &          49.4 &          16.9 & \textbf{0.1}&       0.6 \\
& \texttt{simple\_arbiter\_10}                       &  10 &  10 &   \textbf{2.4}&        1603.2 &       818.0 &  \timeout \\
        \midrule
        \multirow{8}{*}{\rotatebox[origin=c]{90}{\parbox{0.9cm}{\centering \toolname{}\\(19.07)} \emph{vs} \parbox{1.2cm}{\centering \bosy{}\\(July '19)}}}
& \texttt{round\_robin\_arbiter\_4}                  &   4 &   4 &           1.3 &           4.1 & \textbf{1.2}&    3178.3 \\
& \texttt{ltl2dba\_Q\_6}                             &   6 &   1 &   \textbf{0.9}&           4.9 &         1.1 &    1539.9 \\
& \texttt{detector\_10}                              &  10 &   1 &   \textbf{0.8}&         486.4 &     \memout &     506.0 \\
& \texttt{ltl2dba\_C2\_10}                           &  10 &   1 &   \textbf{1.1}&         457.2 &     \memout &     474.0 \\
& \texttt{SPIWriteManag}                             &   5 &  15 &   \textbf{3.3}&           3.6 &        16.3 &    1076.3 \\
& \texttt{SPI}                                       &   4 &  16 &   \textbf{5.1}&           6.6 &        10.0 &    1341.1 \\
& \texttt{ltl2dba\_R\_6}                             &   6 &   1 &   \textbf{4.8}&      \timeout &     \memout &    1167.0 \\
& \texttt{OneCounterGuiA6}                           &   9 &   9 &   \textbf{2.9}&           7.3 &        79.0 &     464.7 \\
        \midrule
        \multirow{16}{*}{\rotatebox[origin=c]{90}{\emph{selected parameterised instances}}}
& \texttt{amba\_...\_arbiter\_2}                       &   3 &   4 &           1.3 &           0.9 & \textbf{0.1}&       0.1 \\
& \texttt{amba\_...\_arbiter\_4}                       &   5 &   6 &           1.7 &           1.4 & \textbf{1.0}&       4.8 \\
& \texttt{amba\_...\_arbiter\_6}                       &   7 &   8 &   \textbf{9.4}&          40.1 &        38.3 &     393.1 \\
& \texttt{amba\_...\_arbiter\_8}                       &   9 &  10 & \textbf{177.7}&        1629.1 &    \timeout &  \timeout \\
& \texttt{full\_arbiter\_enc\_6}                       &   3 &   3 &          11.7 &   \textbf{2.0}&        17.2 &     200.0 \\
& \texttt{full\_arbiter\_enc\_8}                       &   3 &   4 &          63.0 &  \textbf{21.4}&    \timeout &  \timeout \\
& \texttt{full\_arbiter\_enc\_10}                      &   4 &   4 &         337.8 & \textbf{303.7}&    \timeout &  \timeout \\
& \texttt{full\_arbiter\_enc\_12}                      &   4 &   4 &\textbf{1274.5}&      \timeout &    \timeout &  \timeout \\
& \texttt{ltl2dba\_beta\_4}                            &   8 &   1 &   \textbf{1.1}&           6.1 &         1.1 &       2.4 \\
& \texttt{ltl2dba\_beta\_6}                            &  12 &   1 &   \textbf{1.5}&         237.0 &        72.5 &     152.5 \\
& \texttt{ltl2dba\_beta\_8}                            &  16 &   1 &  \textbf{19.0}&      \timeout &   \errorout &   \memout \\
& \texttt{ltl2dba\_beta\_10}                           &  20 &   1 & \textbf{352.2}&      \timeout &   \errorout &   \memout \\
& \texttt{ltl2dba\_Q\_6}                               &   6 &   1 &   \textbf{0.9}&           4.9 &         1.1 &     1539.9 \\
& \texttt{ltl2dba\_Q\_8}                               &   8 &   1 &   \textbf{4.2}&         145.3 &       324.9 &   \timeout \\
& \texttt{ltl2dba\_Q\_10}                              &  10 &   1 &  \textbf{47.5}&      \timeout &    \timeout &   \timeout \\
& \texttt{ltl2dba\_Q\_12}                              &  12 &   1 &\textbf{1042.4}&      \timeout &    \timeout &   \timeout \\
        \bottomrule
    \end{tabular}}
    \caption{Time (s) to decide \textbf{Realisability},
        comparing \toolname{} with its previous version, \ltlsynt{} and \bosy{},
        both pairwise on the specifications with the 8 largest differences and on selected parameterised instances.
        We mark timeouts by \timeout, memouts by \memout, and errors by \errorout.
    }\label{tab:experimental-results-comparison-realizability}
\end{table}

\begin{table}
    \setlength{\tabcolsep}{4pt}
    \centering
    \scalebox{0.9}{
    \begin{tabular}{clrrr@{\hskip 1pt}lr@{\hskip 1pt}lr@{\hskip 1pt}lr@{\hskip 1pt}l}
        \toprule
        & & & & \multicolumn{2}{c}{\toolname{}} & \multicolumn{2}{c}{\toolname{}} &
          \multicolumn{2}{c}{\ltlsynt{}} & \multicolumn{2}{c}{\bosy{}} \\
      & Specification & $|\apin|$ & $|\apout|$ & \multicolumn{2}{c}{(19.07)} & \multicolumn{2}{c}{(CAV'18)} & \multicolumn{2}{c}{(2.8.1)} & \multicolumn{2}{c}{(July '19)} \\
        \midrule
        \multirow{8}{*}{\rotatebox[origin=c]{90}{\parbox{0.9cm}{\centering \toolname{}\\(19.07)} \emph{vs} \parbox{1.2cm}{\centering \toolname{}\\(CAV'18)}}}
& \texttt{tictactoe}                                 &   9 &   9 & \textbf{0} & \cmark &        153 & \cmark &    \timeout & &    \timeout & \\
& \texttt{collector\_v2\_4}                          &   4 &   1 &         20 & \cmark &       2060 & \xmark &       8292 & \cmark & \textbf{10} & \cmark \\
& \texttt{ltl2dba\_Q\_6}                             &   6 &   1 & \textbf{114} & \cmark &       9761 & \cmark &      16563 & \cmark &      48591 & \xmark \\
& \texttt{collector\_v4\_7}                          &   7 &   1 & \textbf{57} & \cmark &       3547 & \xmark &    \timeout & &    \timeout & \\
& \texttt{collector\_v2\_5}                          &   5 &   1 & \textbf{24} & \cmark &       1238 & \cmark &    \timeout & &    \timeout & \\
& \texttt{OneCounter}                                &   9 &   9 & \textbf{22} & \cmark &        870 & \cmark &       2534 & \cmark &    \memout & \\
& \texttt{ltl2dba\_C2\_8}                            &   8 &   1 & \textbf{58} & \cmark &       2133 & \xmark &   \timeout & &    \memout & \\
& \texttt{detector\_8}                               &   8 &   1 & \textbf{58} & \cmark &       2133 & \xmark &    \timeout & &        440 & \cmark \\
        \midrule
        \multirow{8}{*}{\rotatebox[origin=c]{90}{\parbox{0.9cm}{\centering \toolname{}\\(19.07)} \emph{vs} \parbox{1.2cm}{\centering \ltlsynt{}\\(2.8.1)}}}
& \texttt{narylatch\_10}                             &  11 &  10 & \textbf{186} & \xmark &    \timeout & &    6349541 & \xmark &    \timeout & \\
& \texttt{simple\_arbiter\_10}                       &  10 &  10 & \textbf{29} & \cmark &    \memout & &     996921 & \xmark &    \timeout & \\
& \texttt{prioritized\_arbiter\_8}                   &   9 &   9 & \textbf{43} & \cmark &    \memout & &     747766 & \xmark &    \memout & \\
& \texttt{ltl2dba\_C2\_6}                            &   6 &   1 & \textbf{31} & \cmark &       1071 & \cmark &     422872 & \xmark &        468 & \cmark \\
& \texttt{detector\_6}                               &   6 &   1 & \textbf{31} & \cmark &       1071 & \cmark &     422872 & \xmark &        449 & \cmark \\
& \texttt{ltl2dba\_E\_10}                            &  10 &   1 & \textbf{29} & \cmark &   \timeout & &     321799 & \xmark &    \timeout & \\
& \texttt{simple\_arbiter\_8}                        &   8 &   8 & \textbf{18} & \cmark &    \memout & &      87834 & \xmark &         21 & \cmark \\
& \texttt{narylatch\_8}                              &   9 &   8 & \textbf{144} & \cmark &    \memout & &     376661 & \xmark &    \timeout & \\
        \midrule
        \multirow{8}{*}{\rotatebox[origin=c]{90}{\parbox{0.9cm}{\centering \toolname{}\\(19.07)} \emph{vs} \parbox{1.2cm}{\centering \bosy{}\\(July '19)}}}
& \texttt{simple\_arbiter\_enc\_6}                   &   3 &   3 & \textbf{751} & \cmark &       2699 & \xmark &       5276 & \xmark &     611824 & \xmark \\
& \texttt{prioritized\_arbiter\_enc\_6}              &   3 &   3 & \textbf{1445} & \xmark &       7938 & \xmark &      66443 & \xmark &     897716 & \xmark \\
& \texttt{ltl2dba\_Q\_6}                             &   6 &   1 & \textbf{114} & \cmark &       9761 & \cmark &      16563 & \cmark &      48591 & \xmark \\
& \texttt{full\_arbiter\_7}                          &   7 &   7 & \textbf{911} & \xmark &    \memout & &     112488 & \xmark &     388240 & \xmark \\
& \texttt{ltl2dba\_E\_6}                             &   6 &   1 & \textbf{17} & \cmark & \textbf{17} & \cmark &       3630 & \cmark &       5813 & \cmark \\
& \texttt{full\_arbiter\_6}                          &   6 &   6 & \textbf{373} & \cmark &        634 & \cmark &      28264 & \xmark &     117163 & \xmark \\
& \texttt{round\_robin\_arbiter\_4}                  &   4 &   4 & \textbf{94} & \cmark &        239 & \cmark &       1928 & \cmark &      19738 & \cmark \\
& \texttt{full\_arbiter\_5}                          &   5 &   5 & \textbf{284} & \cmark &        351 & \cmark &       7039 & \cmark &      47570 & \xmark \\
        \bottomrule
    \end{tabular}}
    \caption{Size of AIG for \textbf{Synthesis} and \textbf{Quality},
        comparing \toolname{} with its previous version, \ltlsynt{} and \bosy{},
        pairwise on the specifications with the 8 largest differences.
        We mark timeouts by \timeout, memouts by \memout, and errors by \errorout.
        The symbol \cmark{} denotes successful verification and \xmark{} denotes a timeout during verification of the AIG.
    }\label{tab:experimental-results-comparison-synthesis}
\end{table}

\begin{table}[tbp]
    \setlength{\tabcolsep}{2pt}
    \renewcommand*{\arraystretch}{1.5}
    \centering
    \footnotesize
    \begin{tabular}{lrrrrc@{\hskip 10pt}rrrr}
\toprule
& \multicolumn{4}{c}{Time for realisability}
&
& \multicolumn{4}{c}{Size of AIG} \\
\cmidrule{2-5}
\cmidrule{7-10}
& \parbox{0.9cm}{\centering \toolname{}\\(19.07)}
& \parbox{1.2cm}{\centering \toolname{}\\(CAV'18)}
& \parbox{1.2cm}{\centering \ltlsynt{}\\(2.8.1)}
& \parbox{1.2cm}{\centering \bosy{}\\(July '19)}
&
& \parbox{0.9cm}{\centering \toolname{}\\(19.07)}
& \parbox{1.2cm}{\centering \toolname{}\\(CAV'18)}
& \parbox{1.2cm}{\centering \ltlsynt{}\\(2.8.1)}
& \parbox{1.2cm}{\centering \bosy{}\\(July '19)}
\\
\midrule
\parbox{1.2cm}{\centering \toolname{}\\(19.07)}
& --- &   105 &   124 &   145 &
& --- &   212 &   293 &   190 \\[0.7em]
\parbox{1.2cm}{\centering \toolname{}\\(CAV'18)}
&    25 & --- &    80 &    98 &
&    31 & --- &   242 &   125 \\[0.7em]
\parbox{1.2cm}{\centering \ltlsynt{}\\(2.8.1)}
&    22 &    34 & --- &    77 &
&     0 &    23 & --- &    82 \\[0.7em]
\parbox{1.2cm}{\centering \bosy{}\\(July '19)}
&    14 &    33 &    32 & --- &
&   101 &   139 &   202 & --- \\
\bottomrule
    \end{tabular}
    \caption{Cross-comparison of different tool. Each cell counts the number of instances where the result of the
        row tool is strictly better than the result of the column tool, comparing time for realisability and size of the AIG
        as the number of and gates plus number of latches.
        A time is only considered better if it is at least 5\;seconds less than the other time.
}\label{tab:experimental-results-comparison-cross}
\end{table}

The results for these comparisons are given in table
Table~\ref{tab:experimental-results-comparison-realizability} for time to decide realisability, in
Table~\ref{tab:experimental-results-comparison-synthesis} for size of the solutions, and
Table~\ref{tab:experimental-results-comparison-cross} gives a cross-comparison.

\paragraph{Analysis.}

Compared to the previous version of \toolname{} we solve at least 40 additional specifications in the \textbf{Realisability} and \textbf{Synthesis} categories and see a considerable improvement in the \textbf{Quality} ratings.
We believe that this is partly due to the symbolic construction and representation of the transition relation which applies to the arena and the automata for LTL fragments, e.g. safety.
Only on smaller instances of the \texttt{full\_arbiter} specification the previous version is faster. This might be related to the overhead of using  symbolic data-structures for small alphabets.
Further, the revised AIG encoding strategy yields smaller controllers in comparison to the previous version.

Compared to \bosy{}, our approach can scale better on larger and complex specifications.
One can observe this on parameterised specifications that mainly increase the number of input propositions.
Even though \bosy{} employs an input-symbolic QBF encoding it could not deal with the large specifications. We hypothesise that this is caused by an explicit representation somewhere in the synthesis chain.
For synthesis, while \bosy{} produces a smaller solution in 101 cases, \toolname{} produces a smaller solution in 129 cases, and often by a much larger factor.
We believe this is due to our structured encoding, which is hard to recover from a solution given by the constraint solver that bounded synthesis employs.

\subsection{\emph{RQ2}: Comparison of different exploration strategies}\label{subsec:comp-expl}

\begin{table}[tbp]
    \setlength{\tabcolsep}{2pt}
    \renewcommand*{\arraystretch}{1.5}
    \centering
    \footnotesize
    \begin{tabular}{lrrrrr}
\toprule
$|\venv|$ & $\oexplbfs$ & $\oexplbfsclear$ & $\oexplpq$ & $\oexplpqclear$ & $\oexpl^{\textsf{best}}$ \\
\midrule
$\oexplbfs$
& --- &    10 &    45 &    36 & 5 \\
$\oexplbfsclear$
&    30 & --- &    50 &    37 & 11 \\
$\oexplpq$
&    31 &    28 & --- &     6 & 5 \\
$\oexplpqclear$
&    45 &    45 &    48 & --- & 30 \\
\bottomrule
    \end{tabular}\qquad\quad
    \begin{tabular}{lrrrrr}
\toprule
Time & $\oexplbfs$ & $\oexplbfsclear$ & $\oexplpq$ & $\oexplpqclear$ & $\oexpl^{\textsf{best}}$ \\
\midrule
$\oexplbfs$
& --- &    14 &    57 &    30 &   1 \\
$\oexplbfsclear$
&    43 & --- &    62 &    45 &  31 \\
$\oexplpq$
&    13 &     9 & --- &     2 &   1 \\
$\oexplpqclear$
&    32 &    22 &    47 & --- &  13 \\
\bottomrule
    \end{tabular}
    \caption{Cross-comparison of different exploration strategies. Each cell counts the number of instances where the result of the
        row strategy is strictly better than the result of the column strategy. $\oexpl^{\textsf{best}}$ is the best of the three other exploration strategies.
    A time is only considered better if it is at least 5\;seconds less than the other time.
}\label{tab:experimental-results-exploration-cross}
\end{table}

\begin{table}[tbp]
    \setlength{\tabcolsep}{2pt}
    \centering
    \scriptsize
    \begin{tabular}[t]{lrrrrrrrr}
        \toprule
        &
        \multicolumn{4}{c}{Explored states $|\venv|$} &
        \multicolumn{4}{c}{Time for realisability (s)} \\
        \cmidrule(r){2-5} \cmidrule(r){6-9}
        Specification &
        $\oexplbfs$ & $\oexplbfsclear$ & $\oexplpq$ & $\oexplpqclear$ &
        $\oexplbfs$ & $\oexplbfsclear$ & $\oexplpq$ & $\oexplpqclear$ \\
        \midrule
\texttt{amba\_case\_study\_2}                      &       8187 &       3852 &       5598 & \textbf{2428} &    1008.9 &     451.4 &     707.3 &     \textbf{306.8} \\
\texttt{collector\_v3\_5}                          &         78 &         73 &        148 & \textbf{54} &       \textbf{2.3} &       \textbf{2.3} &       4.3 &       2.6 \\
\texttt{collector\_v3\_6}                          &        148 &        147 &        206 & \textbf{32} &       6.5 &       6.0 &      13.0 &       \textbf{4.6} \\
\texttt{collector\_v3\_7}                          &        306 &        305 &        386 & \textbf{22} &      42.2 &      39.9 &      94.0 &      \textbf{18.2} \\
\texttt{detector\_unreal\_12}                      &         28 &         31 &         48 & \textbf{22} &       \textbf{1.7} &       1.8 &       2.9 &       2.1 \\
\texttt{full\_arbiter\_unreal1\_3\_6}              &      65536 &      65536 &       7475 & \textbf{6645} &      23.1 &      \textbf{12.4} &      30.7 &      14.1 \\
\texttt{full\_arbiter\_unreal1\_3\_8}              &    1685603 &    1678429 &       6576 & \textbf{6321} &     369.9 &     328.4 &      34.4 &      \textbf{17.3} \\
\texttt{full\_arbiter\_unreal1\_3\_10}             &        --- &        --- &       6812 & \textbf{6249} &   \timeout &   \timeout &      38.0 &      \textbf{16.7} \\
\texttt{full\_arbiter\_unreal1\_3\_12}             &        --- &        --- &       6899 & \textbf{6418} &   \timeout &   \timeout &      53.1 &      \textbf{19.4} \\
\texttt{genbuf2}                                   &      77479 &      61878 &      83025 & \textbf{35196} &      80.1 &      \textbf{56.9} &     175.9 &      77.5 \\
\texttt{KitchenTimerV2}                            &         23 & \textbf{12} &         24 &         24 &       2.7 &       \textbf{2.1} &       3.2 &       3.1 \\
\texttt{load\_balancer\_unreal1\_2\_10}            &       2523 & \textbf{572} &        701 &        700 &      12.2 &       2.6 &       2.3 &       \textbf{1.5} \\
\texttt{load\_balancer\_unreal1\_2\_12}            &      10106 & \textbf{638} &        725 &        713 &      22.4 &       2.6 &       2.3 &       \textbf{1.9} \\
\texttt{ltl2dba\_R\_6}                             &        162 &         80 &         29 & \textbf{14} &       4.7 &       4.0 &       3.8 &       \textbf{3.0} \\
\texttt{ltl2dba\_R\_8}                             &        138 &        138 &          4 & \textbf{3} &     603.1 &     613.2 &     544.7 &     \textbf{437.6} \\
\texttt{round\_robin\_[...]\_unreal1\_2\_12}       &      11703 &      16384 & \textbf{4096} & \textbf{4096} &       \textbf{8.1} &      15.0 &      22.9 &      13.0 \\
\texttt{round\_robin\_[...]\_unreal1\_2\_15}       &      93623 &      93623 &      22509 & \textbf{21077} &      \textbf{21.7} &      40.2 &      66.6 &      25.2 \\
\texttt{round\_robin\_[...]\_unreal1\_2\_18}       &     673158 &     669697 &     262144 & \textbf{45152} &     140.5 &     117.8 &     243.9 &      \textbf{57.7} \\
\texttt{SliderDelayed}                             &         81 & \textbf{48} &         81 &         81 &       1.5 &       \textbf{1.0} &       2.0 &       1.7 \\
\texttt{slider\_delayed}                           & \textbf{50} &         81 &         81 &         81 &       \textbf{1.3} &       1.4 &       2.2 &       1.4 \\
\texttt{TwoCounters3}                              & \textbf{11} & \textbf{11} &         99 &         99 &       9.1 &       \textbf{8.2} &      44.5 &      40.4 \\
\texttt{TwoCountersInRangeM0}                      & \textbf{7} & \textbf{7} &         20 &         19 &       2.9 &       \textbf{2.3} &       7.8 &       6.7 \\
\texttt{TwoCountersInRangeM1}                      & \textbf{7} & \textbf{7} &         24 &         23 &       2.7 &       \textbf{2.4} &       9.1 &       7.9 \\
        \bottomrule
    \end{tabular}
    \caption{Comparison of different exploration strategies for checking realisability.
    }\label{tab:experimental-results-exploration}
\end{table}

\paragraph{Experimental Design.}

We compare the four different exploration strategies $\oexplbfs$, $\oexplbfsclear$, $\oexplpq$ and $\oexplpqclear$ and measure the number of explored states and time needed to check realisability. Table~\ref{tab:experimental-results-exploration-cross} gives a cross-comparison in order to identify a dominant approach. Further, Table~\ref{tab:experimental-results-exploration} lists runtimes and number of explored states where there is a significant difference in the number of explored states. 

\paragraph{Analysis.}

Table~\ref{tab:experimental-results-exploration-cross} suggests that $\oexplpqclear$ can avoid the exploration of states not relevant for deciding realisability. However, this efficiency seems to be costly, since in the runtime comparison it falls behind the $\oexplbfsclear$ configuration.

We see that filtering the queue in $\oexplbfsclear$ and $\oexplpqclear$ in comparison to $\oexplbfs$ and $\oexplpq$, respectively, often reduces the amount of explored states. However, filtering the queue generally incurs overhead, as the reachable state space needs to be explored and the queue rebuilt.
For instance, on \texttt{amba\_case\_study\_2}, we half the number of states explored when filtering is used.
Further, comparing $\oexplpq$ with $\oexplbfs$, we see that using the scoring-based exploration can significantly reduce the amount of explored states.
This especially seems to hold for unrealisable specifications. We assume this is because the environment player
only needs to find a path to a state that forces the controller player to violate the specification, while the remaining states only contribute to satisfying the specification. However, $\oexplpq$ and $\oexplpqclear$ can also drive the exploration of the state-space in the wrong direction as seen for example in \texttt{TwoCountersInRangeM0} and \texttt{TwoCounters3}.

An artefact of the in-parallel running parity game construction and parity game solution can be observed for the \texttt{collector\_v3\_[5-7]} and the \texttt{ltl2dba\_R\_[6-8]} specifications. Here the construction of the states is faster than the algorithm solving the game and when the translation slows down with the increasing size of the alphabet the solver catches up.

\subsection{\emph{RQ3}: Comparison of different minimisation strategies}\label{subsec:comp-min}

\begin{table}[tbp]
    \setlength{\tabcolsep}{2pt}
    \renewcommand*{\arraystretch}{1.5}
    \centering
    \footnotesize
    \begin{tabular}{lrrrrr}
    \toprule
$|\text{AIG}|$ & $\lunstruct$ & $\lunstruct^{\textsf{min}}$ & $\lstruct$ & $\lstruct^{\textsf{min}}$ & $l_\textsf{best}$ \\
\midrule
$\lunstruct$
& --- &    27 &   131 &   120 &   18 \\
$\lunstruct^{\textsf{min}}$
&   157 & --- &   172 &   200 &  138 \\
$\lstruct$
&    80 &    61 & --- &    71 &   29 \\
$\lstruct^{\textsf{min}}$
&   106 &    28 &   117 & --- &    3 \\
\bottomrule
    \end{tabular}
    \caption{Cross-comparison of different minimisation strategies. Each cell counts the number of instances where the size of the
    AIG from the row strategy is strictly smaller than the result of the column strategy, and $l_\textsf{best}$ gives the minimum over all three other strategies.
}\label{tab:experimental-results-minimization-cross}
\end{table}

\begin{table}[tbp]
    \setlength{\tabcolsep}{2pt}
    \centering
    \scriptsize
    \begin{tabular}[t]{c@{\hskip 5pt}lrrrr@{\hskip 1pt}lr@{\hskip 1pt}lr@{\hskip 1pt}lr@{\hskip 1pt}l}
        \toprule
        & & & & $\oexplbfs$ & \multicolumn{8}{c}{Size of AIG} \\
        \cmidrule(r){5-5}
        \cmidrule(r){6-13}
        & Specification & $|\apin|$ & $|\apout|$  & $|\venv|$ &
        \multicolumn{2}{c}{$\lunstruct$} & \multicolumn{2}{c}{$\lunstruct^{\textsf{min}}$} & \multicolumn{2}{c}{$\lstruct$} & \multicolumn{2}{c}{$\lstruct^{\textsf{min}}$} \\
        \midrule
& \texttt{collector\_v2\_5}                          &   5 &   1 &      144 &        446 & \cmark & \textbf{24} & \cmark &         73 & \cmark &         35 & \cmark \\
& \texttt{collector\_v2\_6}                          &   6 &   1 &      336 &       1610 & \cmark & \textbf{34} & \cmark &        101 & \cmark &         51 & \cmark \\
& \texttt{collector\_v2\_7}                          &   7 &   1 &      607 &       3052 & \xmark & \textbf{65} & \cmark &        113 & \cmark &         77 & \cmark \\
& \texttt{collector\_v4\_6}                          &   6 &   1 &      184 &        772 & \cmark &       1216 & \cmark & \textbf{209} & \cmark &       2020 & \cmark \\
& \texttt{collector\_v4\_7}                          &   7 &   1 &      375 &       1451 & \cmark & \textbf{57} & \cmark &        141 & \cmark &        173 & \cmark \\
& \texttt{full\_arbiter\_6}                          &   6 &   6 &      645 & \textbf{373} & \cmark &       3865 & \xmark &        758 & \cmark &       8469 & \xmark \\
& \texttt{full\_arbiter\_7}                          &   7 &   7 &     1422 & \textbf{911} & \xmark &      10722 & \xmark &       1080 & \cmark &      20349 & \xmark \\
& \texttt{full\_arbiter\_enc\_8}                     &   3 &   4 &    15985 &      18533 & \xmark &      11623 & \xmark & \textbf{2904} & \xmark &      31195 & \xmark \\
& \texttt{full\_arbiter\_enc\_10}                    &   4 &   4 &    70662 &   \timeout & & \textbf{147730} & \xmark &   \timeout & &   \timeout & \\
& \texttt{KitchenTimerV3}                            &   4 &   6 &       69 &         48 & \cmark & \textbf{1} & \cmark &         64 & \cmark &         14 & \cmark \\
& \texttt{KitchenTimerV4}                            &   4 &   6 &       78 &         48 & \cmark & \textbf{1} & \cmark &         64 & \cmark &         14 & \cmark \\
& \texttt{load\_balancer\_8}                         &   9 &   8 &     2250 &      12903 & \xmark &       3751 & \cmark & \textbf{512} & \cmark &       4705 & \cmark \\
& \texttt{ltl2dba\_beta\_6}                          &  12 &   1 &       49 &       1026 & \cmark &       1026 & \cmark & \textbf{90} & \cmark & \textbf{90} & \cmark \\
& \texttt{ltl2dba\_beta\_8}                          &  16 &   1 &       81 &       1819 & \cmark &       1819 & \cmark & \textbf{151} & \cmark & \textbf{151} & \cmark \\
& \texttt{ltl2dba\_beta\_10}                         &  20 &   1 &      121 &       2445 & \cmark &       2445 & \cmark & \textbf{156} & \cmark & \textbf{156} & \cmark \\
& \texttt{ltl2dba\_Q\_6}                             &   6 &   1 &      170 &       3504 & \cmark &       3504 & \cmark & \textbf{114} & \cmark & \textbf{114} & \cmark \\
& \texttt{ltl2dba\_Q\_8}                             &   8 &   1 &      986 &      31821 & \xmark &      31821 & \xmark & \textbf{187} & \cmark & \textbf{187} & \cmark \\
& \texttt{ltl2dba\_Q\_10}                            &  10 &   1 &     5742 &   \timeout & &     199301 & \xmark & \textbf{243} & \cmark & \textbf{243} & \cmark \\
& \texttt{ltl2dba\_Q\_12}                            &  12 &   1 &    33462 &   \timeout & &   \timeout & & \textbf{309} & \cmark &   \timeout & \\
& \texttt{prioritized\_arbiter\_enc\_8}              &   4 &   4 &    23554 &      57136 & \xmark &      33762 & \xmark & \textbf{3487} & \xmark &      85101 & \xmark \\
& \texttt{prioritized\_arbiter\_enc\_10}             &   4 &   4 &   142337 &   \timeout & &   \timeout & & \textbf{6906} & \xmark &   \timeout & \\
& \texttt{prioritized\_arbiter\_enc\_12}             &   4 &   4 &   593921 &   \timeout & &    \memout & & \textbf{12256} & \xmark &  \memout & \\
& \texttt{round\_robin\_arbiter\_7}                  &   7 &   7 &     9189 &      12119 & \xmark & \textbf{843} & \cmark &       2535 & \cmark &       1183 & \cmark \\
& \texttt{simple\_arbiter\_enc\_8}                   &   3 &   4 &    10240 &      15906 & \xmark &      11308 & \xmark & \textbf{1317} & \xmark &      24468 & \xmark \\
& \texttt{simple\_arbiter\_enc\_10}                  &   4 &   4 &    65024 &     112422 & \xmark &      62369 & \xmark & \textbf{3356} & \xmark &   \timeout & \\
& \texttt{TorcsSteeringSmart}                        &   4 &   6 &      421 &       4963 & \xmark & \textbf{67} & \cmark &       5131 & \xmark &        104 & \cmark \\
& \texttt{torcs\_steering\_smart}                    &   4 &   6 &      421 &       5006 & \xmark & \textbf{67} & \cmark &       5210 & \xmark &        104 & \cmark \\
        \midrule
        & \multicolumn{3}{l}{Total solved and verified (unique)} & & 292\,(0) & & 299\,(2) & & 300\,(3) & & 298\,(0) & \\
        \bottomrule
    \end{tabular}
    \caption{Comparison for different minimisation strategies, giving the size of the AIG
        as the number of and gates plus number of latches.
        The symbol \cmark{} denotes successful verification and \xmark{} denotes a timeout during verification of the AIG.
    }\label{tab:experimental-results-minimization}
\end{table}

\paragraph{Experimental Design.}

We compare the effect of Mealy machine minimisation and structured encoding on the size of the resulting AIG.
For this experiment we use the exploration strategy $\oexplbfs$.
We compare the four possible combinations: the unstructured and structured encoding functions applied to the unminimised Mealy machine ($\lunstruct$ and $\lstruct$) and to the minimised Mealy machine ($\lunstruct^{\textsf{min}}$ and $\lstruct^{\textsf{min}}$).
Table~\ref{tab:experimental-results-minimization-cross} gives a cross-comparison over all realisable specifications and Table~\ref{tab:experimental-results-minimization} contains outliers where the different techniques have the largest effect and at the bottom cumulative results for the whole set. 

\paragraph{Analysis.}

The cross-comparison of Table~\ref{tab:experimental-results-minimization-cross} suggests that $\lunstruct^{\textsf{min}}$ is a good all-round strategy, since it yields smaller circuits compared to the three other approaches. However, Table~\ref{tab:experimental-results-minimization} shows that the structured encoding $\lstruct$ can sometimes give a significant
reduction in size, e.g. for \texttt{ltl2dba\_Q\_[6-12]}.
But sometimes there is also an increase in size, e.g. for \texttt{round\_robin\_arbiter\_7}.
This usually happens when minimisation is very effective.
The combination $\lstruct^{\textsf{min}}$ does usually not give an improvement over both $\lstruct$ and $\lunstruct^{\textsf{min}}$ on their own.
For \texttt{ltl2dba\_beta\_[6-10]} we have that it can be implemented by a controller using a vector of bits to remember which combinations of inputs have been encountered. The unstructured encoding does not exploit this fact, but this natural structure is restored by $\lstruct$.
For \texttt{full\_arbiter\_7}, $\lstruct$ gives a larger AIG than $\lunstruct$, however it is easier
to verify for our model checker, possibly due to the structure kept by $\lstruct$. Due to these different characteristics we think that a portfolio approach is a sensible default configuration for \toolname{}.

\section{Conclusion and Future Work}\label{sec:fut}

The success of the described approach implemented in \toolname{} relies on several key factors: (1) a demand-driven construction of the automata and the corresponding arena; (2) LTL translations that produce small deterministic automata on-the-fly; (3) a strategy iteration algorithm for solving parity games; especially the fact that the computed optimal strategy in an exploration step serves as a good initial strategy when computing the strategy for the next step; (4) semantic information that can be used for exploration guidance and controller extraction.

While the experimental evaluation places \toolname{} ahead of other competing tools, specifications with large alphabets are still a challenge and need to be addressed. We think restricting the automata constructions of a decomposed formula to letters not violating other parts of the formula, e.g., safety conditions, could be beneficial, as already shown in \cite{DBLP:journals/sttt/SohailS13}. Further, we think we only have scratched the surface of what can be done using the available semantic information and we want to explore potential applications for guiding the exploration of the arena and for extracting implementations, such as circuits but also reactive programs. There are already attempts to extract reactive programs using bounded synthesis \cite{DBLP:conf/atva/GerstackerKF18}, but unfortunately this only works for toy examples. Work in this area could help addressing the third challenge  mentioned in the introduction (representation of synthesised implementation).

\begin{acknowledgements}
We want to thank the anonymous reviewers for their helpful comments and remarks on this manuscript. We also want to thank Swen Jacobs and Guillermo A. Pérez for valuable feedback and testing of \toolname{} during \SC{}.
\end{acknowledgements}

\bibliography{ref}{}       % name your BibTeX data base
%\bibliographystyle{spbasic} % basic style, author-year citations
\bibliographystyle{spmpsci}  % mathematics and physical sciences

\end{document}